\documentclass[%
 reprint,
 superscriptaddress,
 nofootinbib,
 amsmath,amssymb,
 aps,
]{revtex4-2}

\usepackage{graphicx}
\usepackage{dcolumn}
\usepackage{bm}
\usepackage{orcidlink}
\usepackage{hyperref}
\hypersetup{
  pdfborder={0 0 0},
  colorlinks=false
}
\usepackage{array}
\newcolumntype{L}[1]{>{\raggedright\arraybackslash}p{#1}} 
\newcolumntype{C}[1]{>{\centering\arraybackslash}p{#1}}   

\usepackage[dvipsnames]{xcolor} 
\usepackage{subfigure, subcaption}

\usepackage{import}

\newcommand{\etaInf}{\eta^{\bowtie}}
\newcommand{\etaGeo}{\eta^{\triangle}}
\newcommand{\etaCom}{\eta^{\dagger}}

\begin{document}

\preprint{APS/123-QED}

\title{Generalising thermodynamic efficiency of interactions: \\ inferential, information-geometric and computational perspectives}

\author{Qianyang Chen\orcidlink{0000-0001-5382-1396}}
\thanks{Corresponding author: qianyang.chen@sydney.edu.au}
\affiliation{Centre for Complex Systems, Faculty of Engineering, The University of Sydney, Sydney, NSW 2006, Australia}

\author{Nihat Ay\orcidlink{0000-0002-8527-2579}}
\affiliation{Institute for Data Science Foundations, Hamburg University of Technology, 21073 Hamburg, Germany}
\affiliation{Santa Fe Institute, Santa Fe, NM 87501, USA}

\author{Mikhail Prokopenko\orcidlink{0000-0002-4215-0344}}
\affiliation{Centre for Complex Systems, Faculty of Engineering, The University of Sydney, Sydney, NSW 2006, Australia}

\date{\today}

\begin{abstract}
Self-organizing systems consume energy to generate internal order. The concept of thermodynamic efficiency, drawing from statistical physics and information theory, has previously been proposed to characterise a change in control parameter by relating the resulting predictability gain to the required amount of work. However, previous studies have taken a \textit{system-centric} perspective and considered only single control parameters. Here, we generalise thermodynamic efficiency to multiple control parameters and extend the definition of thermodynamic efficiency to protocols in arbitrary directions, by introducing \textit{directional efficiency}. Taking an \textit{observer-centric} perspective, we derive two novel formulations. The first, an \textit{inferential form}, relates efficiency to fluctuations of macroscopic observables, interpreting thermodynamic efficiency in terms of how well the system parameters can be inferred from observable macroscopic behaviour. The second, an \textit{information-geometric form}, expresses efficiency in terms of the Fisher information matrix, interpreting it with respect to how difficult it is to navigate the statistical manifold defined by the control protocol. This observer-centric perspective is contrasted with the existing system-centric view, where efficiency is considered an intrinsic property of the system. 
\end{abstract} 

\maketitle

\section{\label{sec:intro} Introduction}
Collective systems, such as groups of interacting particles, flocks of birds, schools of fish, or active matter, create and maintain internal order by consuming energy and performing work \cite{ramaswamy2010Mechanics, marchetti2013Hydrodynamics, vicsek2012Collective,crosato2019irreversibility}. Their large-scale complex patterns or coordinated movements arise from simple local interactions, a phenomenon known as self-organization. For example, flocking behaviour emerges as each bird tries to follow its neighbours while avoiding collision. The efficiency of local interactions can be measured by \textit{thermodynamic efficiency}: the ratio of entropy reduction or predictability gain to the generalised work required to vary a system's control parameter~\cite{crosato2018Critical,crosato2018Thermodynamics,harding2018Thermodynamic,nigmatullin2021Thermodynamic,chen2025Why}. 

Thermodynamic efficiency has been studied for various complex dynamical systems, including canonical magnetisation models, such as the Curie-Weiss model~\cite{nigmatullin2021Thermodynamic} and the Ising model~\cite{chen2025Why}, self-propelled particles~\cite{crosato2018Thermodynamics}, urban dynamics~\cite{crosato2018Critical}, and contagion networks~\cite{harding2018Thermodynamic}. Each study examined the thermodynamic efficiency along a single protocol, focusing on a control parameter driving the system across distinct macroscopic states (i.e., phase transition). For example, in canonical magnetisation models, phase transitions were induced by changing parameters such as temperature, external magnetic field strength, and coupling strength~\cite{nigmatullin2021Thermodynamic,chen2025Why}. In modelling the motion of self-propelled particles, increasing the particle alignment strength was observed to drive collective motion from disorder to ordered states~\cite{crosato2018Thermodynamics}. In urban systems, modifying the social disposition parameter, a factor that prioritises the attractiveness of suburbs relative to travel costs, shifts the city layouts between monocentric (single mega-suburb) and polycentric (multiple affluent suburbs) configurations~\cite{crosato2018Critical}. In modelling an epidemic spread, the infection transmission rate controls the transition between non-epidemic and epidemic phases~\cite{harding2018Thermodynamic}. These studies observed that thermodynamic efficiency peaks or diverges at a phase transition, and interpreted this in terms of the system's ability to increase its intrinsic cohesiveness relative to the work carried out near critical regimes. In other words, these approaches adopted a \textit{system-centric view} on the efficiency of collective interactions during self-organization. 

Despite different contexts, these studies demonstrated that self-organizing systems are most energetically efficient in increasing internal order (i.e., predictability) when poised at the critical regimes. Building on these results, recent work \cite{chen2025Why} proposed the thermodynamic efficiency as an intrinsic utility for self-organization and introduced the \textit{principle of super-efficiency}. The principle postulates that collective systems self-organize to criticality because it is the regime where (a) given the amount of work available to change the control parameter, the gain in predictability is maximised, or (b) given the desired predictability gain, the required work is minimised. The study also proposed extending thermodynamic efficiency to systems with multiple control parameters; however, formulating this extension remained an open challenge.

Moreover, the intrinsic utility interpretation treated thermodynamic efficiency as the system's property --- independent of how it is observed and measured. However, the emergence of order in collective systems is inherently linked to the specifically chosen observables. Our choice of order parameters, control parameters, or the coordinate representations used to encode the parameter space can all affect the computed value of thermodynamic efficiency. Yet, no formal framework exists to capture this \textit{observer-centric} perspective, even for the single parameter case.

This paper aims to generalise thermodynamic efficiency to settings with multiple control parameters. We derive two new equivalent formulations and offer different interpretations: one expresses efficiency in terms of the fluctuations of macroscopic observables, interpreted from an inferential perspective; the other in terms of Fisher information, interpreted from the information-geometric perspective. These two interpretations are observer-centric, and we contrast them with the previously developed computational perspective~\cite{crosato2018Thermodynamics,chen2025Why}, which is system-centric.

This paper is structured as follows. Section \ref{sec:eta} reviews the definition of thermodynamic efficiency for a single control parameter and extends the definition for multi-parameter systems explored using arbitrary control protocols, by introducing \textit{directional efficiency}. Sections \ref{sec:fluctuation} and \ref{sec:fisher} extend the concept to multi-parameter settings, introducing two new formulations:  \textit{inferential} and \textit{information-geometric}. In addition, these sections offer observer-centric interpretations of these two forms. Section \ref{sec:example} illustrates the idea through simulations of the 2D Ising model. Section \ref{sec:discussion} concludes with a discussion of the key findings and their implications.

\section{\label{sec:eta} Defining thermodynamic efficiency}
Thermodynamic efficiency \cite{crosato2018Critical,crosato2018Thermodynamics,harding2018Thermodynamic,nigmatullin2021Thermodynamic} quantifies how efficiently a collective system self-organizes. It is a function of the control parameter $\lambda$, which modulates the interactions among the individuals within the collective, and hence changes the system's overall predictability. For example, in the Ising model --- a simple system of interacting binary spins on a lattice --- we may consider the coupling strength between neighbouring spins as a control parameter, and increasing the coupling strength increases the spins' tendency to align, hence bringing order to the system. Thermodynamic efficiency $\eta(\lambda)$ measures how much work done by tuning $\lambda$ translates to increased order. The thermodynamic efficiency of interaction connects deeply to information theory under Jaynes' generalised statistical mechanics framework \cite{jaynes1957Information}.

For a collective system governed by a set of control parameters (or generalised forces) $\{\lambda_i\}$, each corresponding to a conjugate observable (or collective variable) $X_i$ \cite{crooks2007Measuring}, if the system is subject to constraints given by the expected values of the observables $\langle X_i \rangle$, the probability distribution the system's microstate $x$ is given by the maximum entropy distribution:
\begin{equation}
    p(x|\underline{\lambda}) = \frac{1}{Z}e^{-\sum_i \lambda_i\,X_i(x)}
\end{equation}
where the sum $\sum_i \lambda_i\,X_i(x)$ can be viewed as generalised total energy. 

For example, in the canonical square-lattice Ising model, the generalised forces $\{\lambda_i\}=\{-\beta B, -\beta J\}$, with $\beta=\frac{1}{k_BT}$ being the inverse temperature, $B$ the external magnetic field, and $J$ the coupling strength. The corresponding conjugate observables are $\{X_i(x)\} = \{\sum_n \sigma_n, \sum\limits_{<m,n>} \sigma_m\sigma_n\}$, where $\sigma_n$ is the spin of the $n^{th}$ site, and the observables represent the total magnetization and the total pairwise interactions, respectively. 

The partition function $Z$ is:
\begin{equation}
    Z(\underline{\lambda}) = \sum_x e^{-\sum_i \lambda_i\,X_i(x)}
\end{equation}

Under information theory, Shannon entropy $S$ quantifies the uncertainty of a probability distribution $p(x)$ and is defined as \cite{shannon1948Mathematical}:
\begin{equation} \label{eq:shannon_entropy}
    S = -\sum_x p(x) \log p(x)
\end{equation}

In this generalised framework, $\log$ of the partition function, or the Massieu potential $\Psi$ (analogous to the Gibbs potential in thermodynamics), is the difference between the Shannon entropy and the generalised total energy \cite{crooks2007Measuring}:
\begin{equation} \label{eq:massieu_potential}
    \Psi = \log Z = -\sum_{i}\lambda_i \langle X_i \rangle + S
\end{equation}
where $\langle X_i \rangle$ is the average of the observable $X_i$.

Thermodynamic efficiency $\eta$ for a given control parameter $\lambda_i$ is defined as the ratio of reduction in Shannon entropy $S$ (or gain in predictability) to the generalised work $\langle \beta\mathbb{W} \rangle$ expended for tuning the parameter ($\mathbb{W}$ refers to thermodynamic work). Mathematically, it can be expressed as \cite{crosato2018Critical,crosato2018Thermodynamics,harding2018Thermodynamic,nigmatullin2021Thermodynamic}:
\begin{equation} \label{eq:eta}
    \eta(\lambda_j) = -\frac{\partial S(\lambda_j)}{\partial \lambda_j}\bigg /\frac{\partial \langle \beta \mathbb{W}\rangle (\lambda_j)}{\partial \lambda_j}
\end{equation}
where partial derivatives are used to emphasise that the definition holds for settings with multiple control parameters.

In a quasi-static process, the infinitesimal work done by the system to change the control parameter is equal to the reduction in Gibbs free energy $d \langle\mathbb{W}\rangle = -d\mathbb{G}$ \cite[Sec.~15]{landau2011Statistical}. Therefore, the thermodynamic efficiency can be expressed as:
\begin{equation} \label{eq:eta-reversible}
    \eta(\lambda_j) = \frac{\partial S(\lambda_j)}{\partial \lambda_j}\bigg /\frac{\partial \beta \mathbb{G}(\lambda_j)}{\partial \lambda_j}
\end{equation}
where $\mathbb{G}$ is the Gibbs potential, so that \cite[Sec.~31]{landau2011Statistical}:
\begin{equation} \label{eq:free-energy}
    \mathbb{G} = -\frac{1}{\beta} \log Z
\end{equation}

Similar to \eqref{eq:eta}, we define generalised directional efficiency evaluated at $\underline{\lambda}=(\lambda_1, \lambda_2, \dots, \lambda_n)$ for the protocol moving in the direction $\underline{v}=(v_1, \dots, v_n)$ as:
\begin{equation} \label{eq:def_eta_dir}
    \eta_{\underline{v}}(\underline{\lambda}) = \frac{-D_{\underline{v}}S(\underline{\lambda})}{D_{\underline{v}}\langle\beta\mathbb{W}\rangle(\underline{\lambda})}
\end{equation}
where the directional derivative of a scalar function $f: \mathbb{R}^n \to \mathbb{R}$ in the direction of $\underline{v}$ at the point $\underline{\lambda}$ is defined as \cite[~pp. 147-148]{loomis1990Advanced}:
\begin{equation}
    D_{\underline{v}}f(\underline{\lambda}) := \lim_{t\to 0} \frac{f(\underline{\lambda} + t\underline{v}) - f(\underline{\lambda})}{t} = \nabla f \cdot \underline{v} 
\end{equation}

For a single-parameter case, the efficiency $\eta(\lambda_j)$ can be interpreted as the ratio of the amount of information gained about the system to the amount of generalised work done to change the control parameter by a small amount $\delta \lambda_j$. We provide interpretations for the multi-parameter case in more detail in Sections \ref{subsec:inferential_efficiency} and \ref{subsec:infogeo_interp}.

\section{\label{sec:fluctuation} Relating efficiency to fluctuation}
The previous studies focus on a single control parameter and its corresponding observable. In the following section, we turn our focus on the inter-dependencies of multiple observables $\{X_i\}$ in response to changing a specific control parameter $\lambda_j$. We aim to reformulate the thermodynamic efficiency $\eta(\lambda_j)$ in terms of the covariance of observables $\{X_i\}$, which is a measure of how the observables fluctuate together, and provide a statistical perspective on the efficiency of self-organization.

\subsection{\label{subsec:eta_in_fluctuation}Reformulating efficiency in terms of fluctuations}
Using well-known results from generalised statistical physics (see Appendix~\ref{ap:stat_mech}), the rate of entropy change with respect to the control parameter $\lambda_j$ can be expressed as:
\begin{equation} \label{eq:numerator}
  \frac{\partial S(\underline{\lambda})}{\partial \lambda_j} = -\sum_{i}\lambda_i \text{Cov}(X_i, X_j)
\end{equation}
This key expression yields the numerator of the thermodynamic efficiency \eqref{eq:eta-reversible} in terms of the covariance across multiple observables. Turning our attention to the denominator, we use the following representation, which is typically derived by combining \eqref{eq:free-energy} and \eqref{eq:mean-deriv} (see Appendix~\ref{ap:stat_mech}):
\begin{eqnarray} \label{eq:denominator}
  \frac{\partial \beta\mathbb{G}(\underline{\lambda})}{\partial \lambda_j} = -\frac{\partial \log Z(\underline{\lambda})}{\partial \lambda_j} = \langle X_j \rangle
\end{eqnarray}
Substituting \eqref{eq:numerator} and \eqref{eq:denominator} into \eqref{eq:eta-reversible}, we can reformulate the thermodynamic efficiency $\eta(\lambda_j)$ as:
\begin{equation} \label{eq:eta-fluctuation-multi}
  \etaInf(\lambda_j) = -\frac{\sum_i \lambda_i \text{Cov}(X_i, X_j)}{\langle X_j \rangle}
\end{equation}
We refer to this as the inferential form of thermodynamic efficiency, denoted by $\etaInf$. For a single control parameter $\lambda$ and conjugate observable $X$, the efficiency can be simplified to:
\begin{equation} \label{eq:eta-fluctuation-single}
  \etaInf(\lambda) = -\frac{\lambda \text{Var}(X)}{\langle X \rangle}
\end{equation}

Equations \eqref{eq:eta-fluctuation-multi} and \eqref{eq:eta-fluctuation-single} offer a key insight: thermodynamic efficiency can be expressed in terms of fluctuations in the observables driven by changing a specific control parameter $\lambda$. Specifically, the efficiency in \eqref{eq:eta-fluctuation-multi} is the total fluctuation, weighted by all the control parameters, and normalised by the mean value of the observable conjugate to the parameter of interest. This formulation links the efficiency of interactions to the statistical inference of the system's probability distribution, and the next section explores this, \textit{observer-centric}, interpretation further.

\subsection{\label{subsec:inferential_efficiency} The statistical perspective of efficiency}
In 1957, Jaynes in his seminal work \cite{jaynes1957Information} posed a key question: What probability distribution best describes our state of knowledge about a physical system, given some average measurement outcomes of macroscopic observables? He concluded that by assigning probabilities which maximise the information entropy subject to relevant constraints, one can arrive at the same probability distribution given by canonical methods of statistical mechanics. The resulting probability distribution is given by $p(x|\underline{\lambda})$, where $\lambda_i$ are the Lagrange multipliers associated with the constraints on the average values of observables $\{X_i\}$.

The same problem can be rephrased differently: Given average measurement values of the observables $\langle X_i(x)\rangle, \langle X_2(x)\rangle...$, how can we infer the parameters of the probability distribution $p(x|\underline{\lambda})$, such that the distribution agrees with the average measurements and maximises the information entropy? This can be thought of as estimating $\underline{\lambda}$ based on sampling the macroscopic observables $X_1(x), X_2(x)...$. The error of estimating $\underline{\lambda}$ from the given data is measured by the covariance matrix of the estimator:
\begin{equation*}
A_{ij} = \text{Cov}(\lambda_i, \lambda_j)
\end{equation*}

The covariance matrix of the estimators $A$ has an inverse relationship with the covariance matrix of the observables $B_{ij} = \text{Cov}(X_i, X_j)$, which measures how much the macroscopic observables $X_i$ and $X_j$ change together \cite[p.~190]{jaynes1963Information}:
\begin{equation*}
  A = B^{-1}
\end{equation*}
This means that when the covariance (and variance) of the macroscopic observables $\text{Cov}(X_i, X_j)$ is large, the covariance (and variance) of the parameter estimators $\text{Cov}(\lambda_i, \lambda_j)$ is small, hence the observer can estimate $\underline{\lambda}$ with smaller errors. 

Therefore, equations \eqref{eq:eta-fluctuation-multi} and \eqref{eq:eta-fluctuation-single} show that the efficiency of self-organization goes hand-in-hand with the \textit{inferential efficiency}. A statistical estimator is considered efficient when it is unbiased and has minimum variance. When the system self-organizes into a state where the observables tend to fluctuate together, it has created internal structure where an observer can recover its hidden parameters more accurately. In other words, when the self-organizing system is efficient, observers can make a better guess of the system's hidden parameters $\{\lambda\}$ by observing the macroscopic observables $\{X\}$. The denominator $\langle X_j \rangle$ normalises the efficiency so that $\eta$ becomes comparable across the set of control parameters $\{\lambda\}$.

\section{\label{sec:fisher} Relating Efficiency to Fisher Information}
\subsection{\label{subsec:eta_in_fisher}Reformulating efficiency in terms of Fisher information}
Thermodynamic efficiency can also be reformulated, using information theory, in terms of the Fisher information \cite{fisher1922Mathematical, cover2005Elements}. As shown in Appendix~\ref{ap:fisher}, $\mathcal{I}_{ij}(\underline{\lambda}) =\text{Cov}(X_i, X_j)$. Substituting this to \eqref{eq:numerator} gives the numerator of the efficiency $\eta(\lambda_j)$ in terms of Fisher information:
\begin{equation} \label{eq:numerator-fisher}
  \frac{\partial S(\underline{\lambda})}{\partial \lambda_j} = -\sum_i \lambda_i \mathcal{I}_{ij}(\underline{\lambda})
\end{equation}

Using \eqref{eq:fisher-covariance} (see Appendix~\ref{ap:fisher}), the denominator of the efficiency $\eta(\lambda_j)$ can be expressed via the integral of the Fisher information $\mathcal{I}_{jj}$ over the control parameter $\lambda_j$ (derivation for single parameter case is given in \cite{crosato2018Thermodynamics, prokopenko2011Relating}): 
\begin{eqnarray}
    \label{eq:denominator-fisher}
  \frac{\partial \beta\mathbb{G}(\underline{\lambda})}{\partial \lambda_j}
  & = & - \frac{\partial \log Z(\underline{\lambda})}{\partial \lambda_j} \notag \\
  & = & - \int_{\lambda_j^*}^{\lambda_j} \mathcal{I}_{jj}(\lambda_1, \dots, \lambda'_{j}, \dots, \lambda_{n}) d\lambda_j'
\end{eqnarray} 
where $\lambda_j^*$ is the point where a change in the control parameter $\lambda_j$ does not extract or perform any work, or the zero-response point. That is:
\begin{equation}
    \frac{\partial \langle\beta\mathbb{W}\rangle (\underline{\lambda})}{\partial \lambda_j} \bigg|_{\lambda_j=\lambda_j^*} = 0
\end{equation}
Under a quasi-static process, this leads to:
\begin{equation}
-\frac{\partial \beta\mathbb{G}(\underline{\lambda})}{\partial \lambda_j} \bigg|_{\lambda_j=\lambda_j^*} = -\langle X_j \rangle= 0
\end{equation}

For example, in a system of interacting spins, such as a 2D Ising model, the zero-response point with respect to the coupling strength is the point where the nearest-neighbour correlations vanish, which is the point where the spin coupling strength is zero.

Combining \eqref{eq:numerator-fisher} and \eqref{eq:denominator-fisher} yields:
\begin{equation} \label{eq:eta-fisher-multi}
  \etaGeo(\lambda_j) = \frac{\sum_i \lambda_i \mathcal{I}_{ij}(\underline{\lambda})}{\int_{\lambda_j^*}^{\lambda_j} \mathcal{I}_{jj}(\lambda_1, \dots, \lambda'_{j}, \dots, \lambda_{n})d\lambda_j'}
\end{equation}
We refer to this as the information-geometric form of thermodynamic efficiency, denoted by $\etaGeo$. For a single control parameter $\lambda$, the efficiency can be simplified to:
\begin{equation} \label{eq:eta-fisher-single}
  \etaGeo(\lambda) = \frac{\lambda \mathcal{I}(\lambda)}{\int_{\lambda^*}^{\lambda} \mathcal{I}(\lambda')d\lambda'}
\end{equation}

The Cramer-Rao inequality states that the covariance matrix of a set of unbiased estimators for the parameters $\{\lambda\}$ is bounded by the inverse of the Fisher information matrix \cite{cover2005Elements}:
\begin{equation} \label{eq:cramer-rao}
    \text{Cov}(\hat{\underline{\lambda}}) \geq \mathcal{I}^{-1}(\underline{\lambda})
\end{equation}
This inequality implies that larger Fisher information leads to a smaller lower bound of the estimation variance, that is, the parameters can be estimated more precisely. In this context, higher thermodynamic efficiency implies the ability to more precisely estimate control parameters $\lambda$ from the macroscopic observables $X$, relative to the total variance accumulated along the control protocol from $\lambda_j$ to $\lambda_j^*$. In other words, continuing along the protocol towards the critical point, a potentially higher precision quantified by the numerator, is contrasted with a potentially higher total precision accumulated along the protocol, quantified by the denominator. This aligns with our earlier point in Section~\ref{subsec:inferential_efficiency} that thermodynamic efficiency is linked to inferential efficiency. However, fundamentally, the equations \eqref{eq:eta-fisher-multi} and \eqref{eq:eta-fisher-single} can be interpreted information-geometrically, as discussed in the next section.

\subsection{\label{subsec:infogeo_interp}Information-geometric perspective of efficiency}
Sections \ref{subsec:eta_in_fluctuation} and \ref{subsec:eta_in_fisher} extended thermodynamic efficiency to multi-parameter systems, but constrain the protocol to follow along the axis of one control parameter at a time. We now examine the thermodynamic efficiency of protocols in arbitrary directions --- referred to as \textit{directional efficiency} (equation \eqref{eq:def_eta_dir}).

The numerator of equation \eqref{eq:def_eta_dir}, which is the directional derivative of entropy reduction in direction $\underline{v}$ evaluated at $\underline{\lambda}$, is given by:
\begin{equation}\label{eq:DvS0}
    -D_{\underline{v}}S(\underline{\lambda}) = -\nabla S(\underline{\lambda}) \cdot \underline{v}
\end{equation}

Substituting each partial derivative $\frac{\partial S}{\partial \lambda_j}$ ($j=1,\dots,n$) in equation \eqref{eq:DvS0} with expression \eqref{eq:numerator-fisher} yields:
\begin{equation}\label{eq:DvS1}
    -D_{\underline{v}}S(\underline{\lambda}) = \sum_j v_j \sum_i \lambda_i\mathcal{I}_{ij}(\underline{\lambda})
\end{equation}

\begin{figure}[ht]
    \centering
    \includegraphics[width=0.7\linewidth]{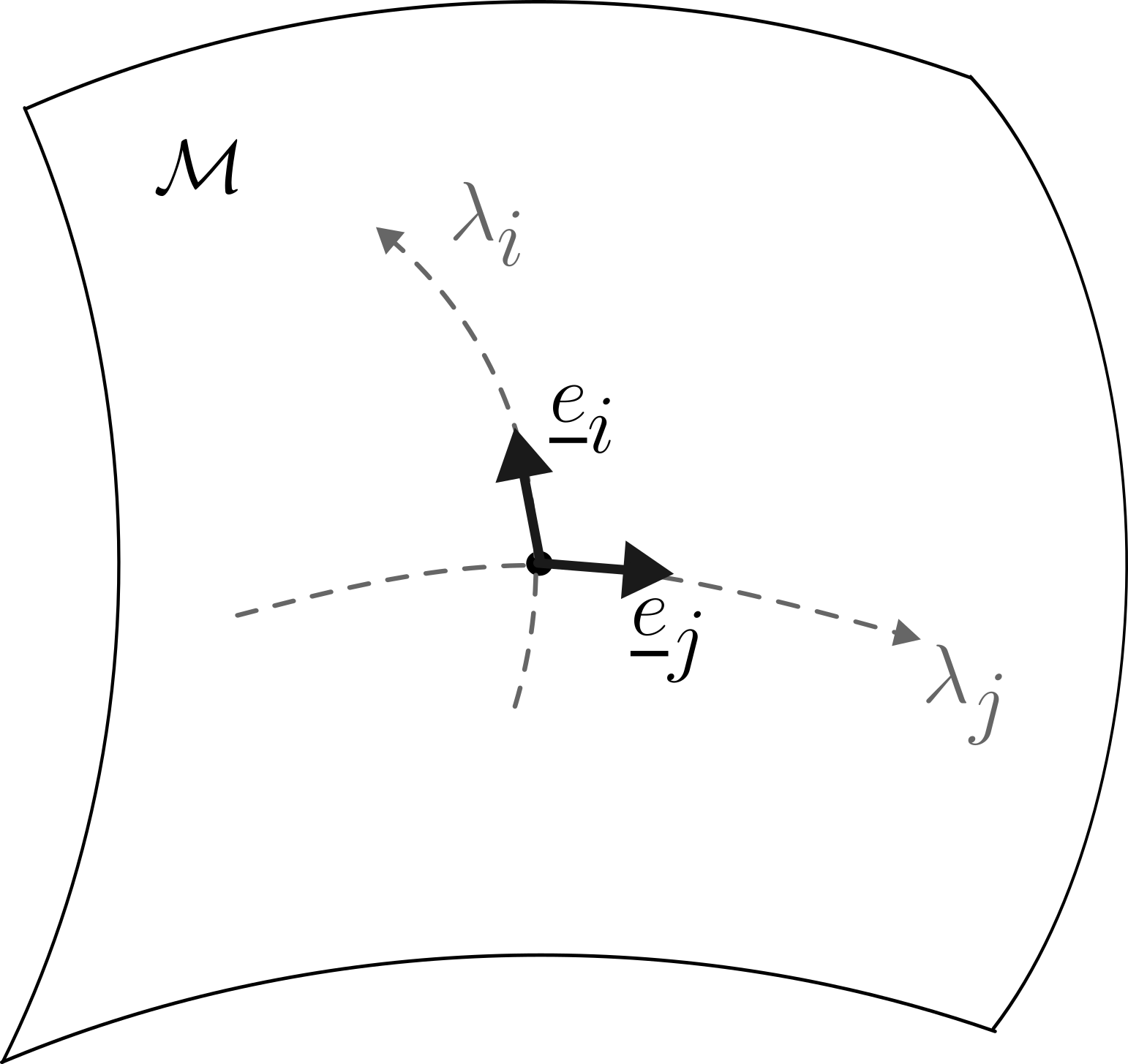}
    \caption{Statistical manifold $\mathcal{M}$. The $\lambda_i$, $\lambda_j$ are coordinate curves, and $\underline{e}_i$, $\underline{e}_j$ are the corresponding tangent vectors. A point on the manifold is a probability density $p(x|\underline{\lambda})$ of a random variable $x$, parameterised by $\underline{\lambda}$.}
    \label{fig:manifold}
\end{figure}

In information geometry, the Fisher information matrix defines a Riemannian metric on a statistical manifold, known as the Fisher-Rao metric. Consider a statistical model $\mathcal{M}= \{p(x|\underline{\lambda})\}$ parameterised by $\underline{\lambda} = (\lambda_1, ..., \lambda_n)$. $\mathcal{M}$ is an m-dimensional manifold with $\underline{\lambda}$ being a coordinate system (illustrated in Figure \ref{fig:manifold}). We define the map from the parameter space to the statistical manifold as $\varphi: \underline{\lambda} \mapsto p(.;\underline{\lambda})$ and the tangent vector $\underline{e}_i$ of the $i^{th}$ coordinate curve on the manifold as:
\begin{equation}
    \underline{e}_i = \frac{\partial \varphi}{\partial \lambda_i}
\end{equation}

A component $\mathcal{I}_{ij}$ of Fisher information matrix is the inner product of the tangent vectors $\underline{e}_i, \underline{e}_j$ of $\mathcal{M}$ in the directions of the coordinate curves $\lambda_i, \lambda_j$ (illustrated in Figure \ref{fig:manifold}) \cite{amari2007Methods, amari2016Information, ay2017Information, amari2021Information}:
\begin{equation} \label{eq:fisher_inner_prod}
    \mathcal{I}_{ij} = \langle \underline{e}_i, \underline{e}_j \rangle
\end{equation}
where $\langle.,.\rangle$ denotes the inner product of two vectors. $\mathcal{I}_{ii}=\langle \underline{e}_i, \underline{e}_i \rangle = \|\underline{e}_i\|^2$ is therefore the squared norm of the tangent vector.

Comparing identity \eqref{eq:fisher_inner_prod} to the numerator of $\eta_{\underline{v}}$ in equation \eqref{eq:DvS1}, we derive that:
\begin{eqnarray} \label{eq:eta_Dv_numerator}
        -D_{\underline{v}}S(\underline{\lambda})
        &=& \sum_j v_j \sum_i \lambda_i\mathcal{I}_{ij}(\underline{\lambda}) \notag\\ 
        &=&\sum_{j}\sum_{i} v_j \lambda_i \langle \underline{e}_i, \underline{e}_j \rangle \notag\\ 
        &=& \langle \underline{v}, \underline{\lambda} \rangle
\end{eqnarray}

Note that equation \eqref{eq:numerator-fisher} written in vector form is:
\begin{equation}
    \nabla S(\underline{\lambda}) = - \mathcal{I}(\underline{\lambda})\underline{\lambda}
\end{equation}

Multiplying both sides of the equation by the inverse of the Fisher information matrix and negative one, we obtain:
\begin{equation}
    -\mathcal{I}^{-1}(\underline{\lambda})\nabla S(\underline{\lambda}) = \underline{\lambda}
\end{equation}
The LHS of the equation is known as the natural gradient of $-S(\underline{\lambda})$, which is the steepest descent direction of $S(\underline{\lambda})$ on the statistical manifold \cite{amari1998natural}.

The vector $\underline{\lambda}$ thus represents both the position on the manifold and the direction of fastest entropy reduction. At the origin, $\underline{\lambda} = (0, \dots, 0)$, the statistical model corresponds to the maximum-entropy (uniform) distribution. The vector $\underline{\lambda}$, which points from the origin to the current position, provides the optimal direction to move away from maximum entropy.

\begin{figure}[ht]
    \centering
    \includegraphics[width=0.7\linewidth]{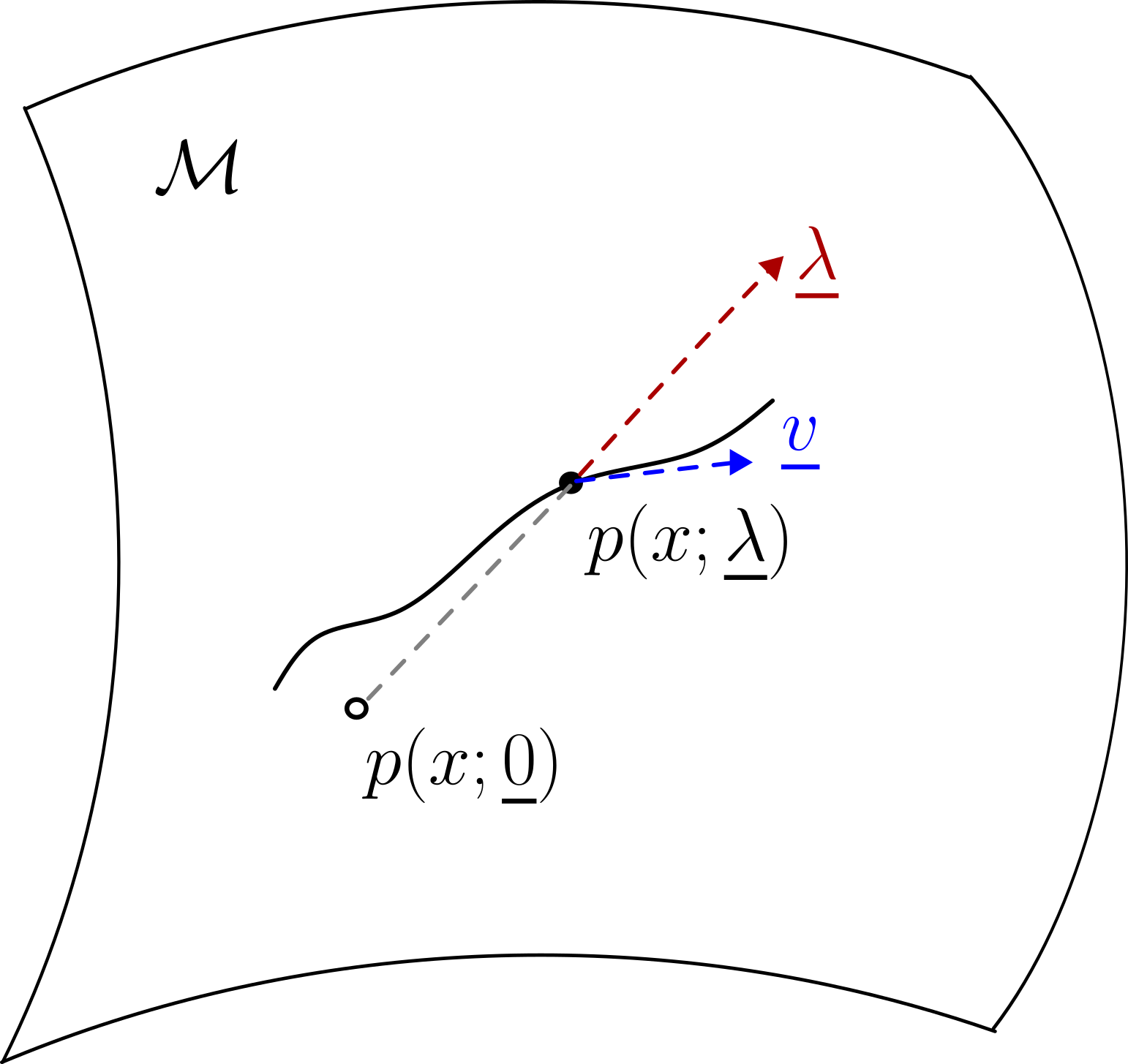}
    \caption{To escape from the maximum-entropy (uniform) distribution $p(x|\underline{0})$, the steepest direction on the statistical manifold is given by the natural gradient of $-S(\underline{\lambda})$ (in red). The numerator of directional efficiency represents the extent to which the protocol's velocity vector (in blue) aligns with this direction.}
    \label{fig:Dv_numerator}
\end{figure}

In this context, the numerator of $\eta_{\underline{v}}$ in equation \eqref{eq:eta_Dv_numerator} can be considered as the inner product of the velocity $\underline{v}$ and the natural gradient of descending entropy evaluated at the point $\underline{\lambda}$. It represents the extent to which the protocol's velocity aligns with the most efficient direction (illustrated in Figure \ref{fig:Dv_numerator}).

On the other hand, the denominator of $\eta_{\underline{v}}$ is the directional derivative of the generalised work. Assuming a quasi-static protocol, the denominator can be expressed as:
\begin{equation} \label{eq:DvW0}
    D_{\underline{v}}\langle\beta\mathbb{W}\rangle (\underline{\lambda}) = D_{\underline{v}}\log Z(\underline{\lambda}) = \nabla \log Z(\underline{\lambda}) \cdot \underline{v}
\end{equation}

Substituting each partial derivative $\frac{\partial \log Z}{\partial \lambda_j}$ in equation \eqref{eq:DvW0} with equation \eqref{eq:denominator-fisher} yields:
\begin{equation} \label{eq:DvW1}
    D_{\underline{v}}\langle\beta\mathbb{W}\rangle (\underline{\lambda})= \sum_j v_j \int_{\lambda_j^*}^{\lambda_j} \mathcal{I}_{jj}(\lambda_1, \dots, \lambda'_j, \dots, \lambda_n) d\lambda'_j
\end{equation}
For each dimension $j=0,\dots,n$, we compute the integral of the corresponding diagonal elements of the Fisher information matrix, starting from the zero-response value $\lambda_j^*$ to the current value $\lambda_j$.

To interpret this equation information-geometrically, we consider the distance and energy of a path on the manifold. The geodesic metric distance for the statistical manifold between two probability distributions $p(x|\underline{\lambda})$ and $p(x|\underline{\lambda}')$, i.e., Fisher-Rao distance~\cite{nielsen2020Elementary}, is defined as: 
\begin{equation}
D_{\text{FR}}(p_{\underline{\lambda}}, p_{\underline{\lambda}'}) := \int_{0}^{1} \sqrt{\dot{\gamma}(t)^{\top} \mathcal{I}_{\gamma(t)}\dot{\gamma}(t)} \;dt 
\end{equation}
where $\gamma$ denotes the geodesic passing through $\gamma(0) = \underline{\lambda}$ and $\gamma(1) = \underline{\lambda}'$. The dot notation represents the time derivative of the curve, i.e., the velocity. 

The energy of the curve $\gamma$ is given by \cite[~p. 20]{jost2017Riemannian}:
\begin{equation} 
E(\gamma) := \frac{1}{2} \int_{0}^{1} \dot{\gamma}(t)^{\top} \mathcal{I}_{\gamma(t)}\dot{\gamma}(t) \;dt 
\end{equation}
$E(\gamma)$ is also known as ``action of the curve" \cite[~p.19]{jost2017Riemannian}, typically defined as the time integral of the difference between kinetic energy and potential energy along the trajectory $\gamma$.

When the system moves in the direction of a coordinate curve $\lambda_j$, the velocity vector $\dot{\gamma}(t)$ is one along axis $\lambda_j$ and zero for all other axes. The energy along the path is therefore simplified to:
\begin{equation} \label{eq:curve_energy}
E(\gamma) = \frac{1}{2} \int_{0}^{1} \mathcal{I}_{jj}(\gamma(t)) \;dt 
\end{equation}

\begin{figure}[ht]
    \centering
    \includegraphics[width=0.7\linewidth]{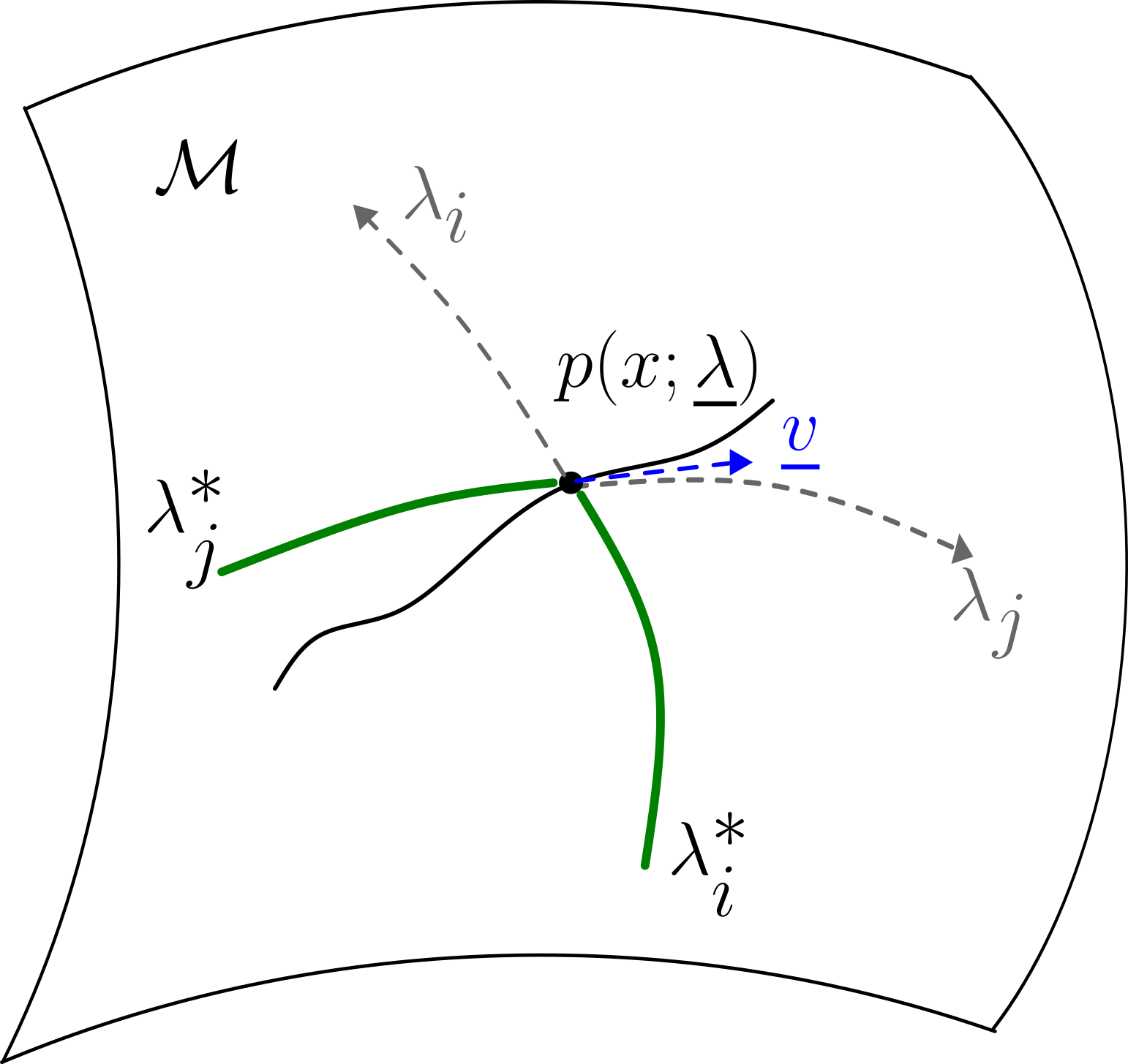}
    \caption{The denominator of directional efficiency represents the weighted sum of the energies of the paths from the zero-response point to the current point over all dimensions. For each dimension $\lambda_j$ ($j=1,2,\dots,n$), the energy $E_j$ of the path (in green) from the zero-response point ($\lambda_j^*$) to its current position ($\lambda_j$) is computed. The energies of these paths are weighted by the velocity of protocol at $\underline{\lambda}$ (in blue).}
    \label{fig:Dv_denominator}
\end{figure}

Comparing equation \eqref{eq:curve_energy} to equation \eqref{eq:DvW1}, and letting $\gamma(0) = \lambda_j^*$ and $\gamma(1)=\lambda_j$, we note that the $j^{th}$ integral in equation \eqref{eq:DvW1} can be considered as the energy of the path going from the reference point $\lambda_j^*$ to the point $\lambda_j$ along the $j^{th}$ axis (subject to a constant scalar multiplier of 2). The denominator of $\eta_{\underline{v}}$ is therefore the sum of these individual path energies across all dimensions, weighted by the control protocol's velocity (Figure \ref{fig:Dv_denominator}). 

Combining equations \eqref{eq:DvS1} and \eqref{eq:DvW1} yields:
\begin{equation} \label{eq:eta_Dv}
    \eta_{\underline{v}}(\underline{\lambda}) = \frac{\langle \underline{v}, \underline{\lambda} \rangle}{\sum_j v_j \int_{\lambda_j^*}^{\lambda_j} \mathcal{I}_{jj}(\lambda_1, \dots, \lambda'_j, \dots, \lambda_n) d\lambda'_j}
\end{equation}

When the protocol is constrained along one axis $\lambda_j$ at the speed of one, equation \eqref{eq:eta_Dv} simplifies to equation \eqref{eq:eta-fisher-multi} in Section \ref{subsec:eta_in_fisher}:
\begin{eqnarray}
    \eta_{(0,\dots, 1, \dots,0)}(\underline{\lambda}) &=& \frac{\sum_k v_k \sum_i \lambda_i\mathcal{I}_{ik}(\underline{\lambda})}{\sum_k v_k \int_{\lambda_k^*}^{\lambda_k} \mathcal{I}_{kk}(\lambda_1, \dots, \lambda'_k, \dots, \lambda_n) d\lambda'_k} \notag \\
    &=& \frac{\sum_i \lambda_i\mathcal{I}_{ij}(\underline{\lambda})}{\int_{\lambda_j^*}^{\lambda_j} \mathcal{I}_{jj}(\lambda_1, \dots, \lambda'_j, \dots, \lambda_n) d\lambda'_j}
\end{eqnarray}
since $v_k=0$ when $k\neq j$ and $v_k=1$ when $k = j$. Thus we recover $\etaGeo(\lambda_j)$ as defined in equation \eqref{eq:eta-fisher-multi}. In this special case, the numerator quantifies the deviation of the protocol's velocity --- restricted along axis $j$ --- from the natural gradient, while the denominator represents the path energy from zero-response point to current position along the $j^{th}$ axis.

Under the information-geometric view, directional efficiency quantifies the efficiency of navigation on the statistical manifold following a control protocol. The directional efficiency is determined by both the system's position ($\underline{\lambda}$) and its velocity ($\underline{v}$) under the protocol. It measures how well the protocol aligns with the optimal direction, given the weighted sum of path energies across all coordinate directions. For a given position, one can compare various protocols by moving the system in different directions $\underline{v}$. Aligning $\underline{v}$ with the natural gradient of the manifold maximises the numerator, allowing the system to gain predictability quickly. However, this may also increase the denominator if $\underline{v}$ places a heavy weight on high-energy axis directions. An efficient protocol must strike a balance between the gain in predictability and the associated energy.

\subsection{Efficiency and parameter space compression}
Previous studies \cite{machta2013Parameter, transtrum2015Perspective} show that physical or biological systems with complex microscopic behaviour can often be effectively described by simple models with a few parameters. The reduction in parameter space dimensionality, or parameter space compression, is possible because the system's behaviour is more sensitive to certain parameter combinations than others. The Fisher information matrix $\mathcal{I}(\underline{\lambda})$ captures this sensitivity: its eigenvalues identify dominant directions in parameter space. A large eigenvalue corresponds to a ``stiff" direction, along which small perturbations to parameter $\lambda_i$ produce large changes in the system's behaviour. 

Thermodynamic efficiency in its Fisher information formulation \eqref{eq:eta-fisher-multi} or \eqref{eq:eta-fisher-single} naturally connects to the concept of parameter space compression. Specifically, $\eta(\lambda_j)$ measures the gain in macroscopic order per unit energetic cost when changing a control parameter $\lambda_j$. If $\lambda_j$ aligns with a stiff direction in the parameter space, then it will have high thermodynamic efficiency because the system responds sharply to it. Conversely, a ``sloppy" parameter direction will contribute very little to macroscopic changes, and so it will have low thermodynamic efficiency.

Therefore, thermodynamic efficiency relates to the compressibility of the parameter space. It is highest along parameter directions that are least compressible --- those that dominate observable behaviours --- and therefore reflects the \textit{hierarchy} of parameter importance (captured by the corresponding eigenvalues of the Fisher Information matrix) with respect to parameter space compression. A self-organizing system that tunes itself to maximise efficiency ``collapses" into its effective parameter space where a few dominating control parameters govern the macroscopic behaviour of the system.

\subsection{Computational interpretation of efficiency}
Considering equations \eqref{eq:eta-fisher-multi} and \eqref{eq:eta-fisher-single}, we can interpret thermodynamic efficiency not only in terms of Fisher information (i.e., in terms of estimation precision), but also in broad computational terms, connecting the efficiency of interactions to distributed computation within the system \cite{crosato2018Thermodynamics}.

It has been suggested in previous studies \cite{crosato2018Thermodynamics,chen2025Why} that the thermodynamic efficiency can be computed in computational terms, for the single parameter case\footnote{We added superscript $\dagger$ to differentiate from other formulations.}:
\begin{equation} \label{eq:eta-computational-single}
    \etaCom(\lambda) = -\frac{d S(\lambda)}{d \lambda}\bigg /\int_{\lambda^*}^{\lambda} \mathcal{I}(\lambda')d\lambda'
\end{equation}
Given our analysis, we can easily generalise this computational interpretation to multiple control parameters:
\begin{equation} \label{eq:eta-computational-multi}
    \etaCom(\lambda_j) = -\frac{\partial S(\underline{\lambda})}{\partial \lambda_j}\bigg /\int_{\lambda_j^*}^{\lambda_j} \mathcal{I}_{jj}(\lambda_1, \dots, \lambda'_{j}, \dots, \lambda_{n})d\lambda_j'
\end{equation}

This computational, \textit{system-centric}, interpretation quantifies the ratio of the predictability gain while computing the next state of the system to the total sensitivity accumulated on a computational path from the reference point (where no work is performed) to the current control parameter value.

\section{\label{sec:example} An example}
To compare the three different forms of thermodynamic efficiency, we consider the canonical 2D Ising model \cite{chen2025ising2D}. The Hamiltonian of the system is given by:
\begin{eqnarray}
\mathcal{H}(\underline{\sigma};J,h) &=& -J\sum_{\langle i,j \rangle} \sigma_i \sigma_j - h\sum_i \sigma_i \notag \\
&=& -JX_1(\underline{\sigma}) - hX_2(\underline{\sigma})
\end{eqnarray}
where $\sigma_i \in \{-1, 1\}$ denotes the spin of the $i^{th}$ site and the macroscopic observables $\{X_1, X_2\}$ capture the nearest-neighbour interactions and total magnetisation respectively. The control parameters considered are the coupling strength $J$ and the external field $h$. 

The inferential form of thermodynamic efficiency is given by \eqref{eq:eta-fluctuation-multi}:
\begin{subequations}
\label{eq:eta-ising-cov}
    \begin{align} 
    \etaInf(J) = -\frac{J \text{Var}(X_1) + h \text{Cov}(X_1, X_2)}{\langle X_1 \rangle}  \\
    \etaInf(h) = -\frac{J \text{Cov}(X_1, X_2) + h \text{Var}(X_2)}{\langle X_2 \rangle} 
    \end{align} 
\end{subequations}
the information-geometric form is given by \eqref{eq:eta-fisher-multi}:  
\begin{subequations}
\label{eq:eta-ising-fisher}
\begin{align} 
    \etaGeo(J) = \frac{J \mathcal{I}_{JJ} + h \mathcal{I}_{Jh}}{\int_{J^*}^{J} \mathcal{I}_{JJ}(J',h)dJ'} \\
    \etaGeo(h) = \frac{J \mathcal{I}_{hJ} + h \mathcal{I}_{hh}}{\int_{h^*}^{h} \mathcal{I}_{hh}(J,h')dh'} 
\end{align} 
\end{subequations} 
and the computational form is given by \eqref{eq:eta-computational-multi}: 
\begin{subequations}
\label{eq:eta-ising-derv}
\begin{align} 
    \etaCom(J) = -\frac{\partial S(J,h)}{\partial J}\bigg /\int_{J^*}^{J} \mathcal{I}_{JJ}(J',h)dJ' \\
    \etaCom(h) = -\frac{\partial S(J,h)}{\partial h}\bigg /\int_{h^*}^{h} \mathcal{I}_{hh}(J,h')dh'
\end{align} 
\end{subequations}  

Zero-response points for $J$ and $h$ can be identified via the average of the corresponding conjugate observables. For a given external field $h$, the zero-response coupling strength $J^*$ is the value of $J$ that yields zero nearest-neighbour correlation, i.e., $\langle X_1\rangle = -\langle \sum \sigma_i\sigma_j\rangle = 0$. For example, when $h=0$,  $J^*=0$. The zero-response point for $h$ is always $h^*=0$ due to symmetry, as the average magnetisation $\langle X_2\rangle = -\langle \sum \sigma_i\rangle=0$ at zero field.

\begin{figure}[!ht]
  \centering
  \includegraphics[width=\linewidth]{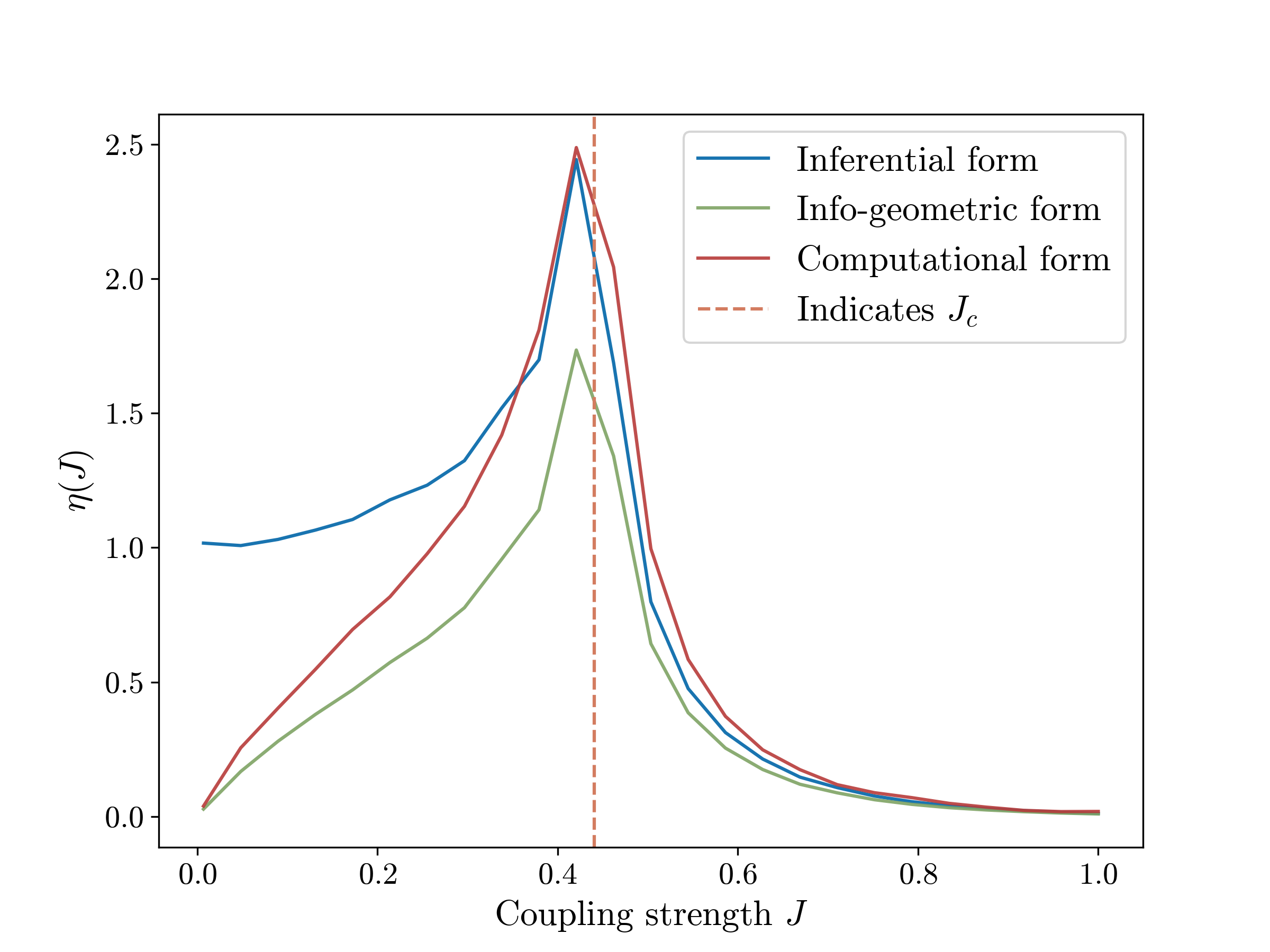}
 \caption{\label{fig:eta_h0} Thermodynamic efficiency computed for the inferential form $\etaInf(J)$, information-geometric form $\etaGeo(J)$, and computational form $\etaCom(J)$, using the canonical 2D Ising model with zero external field.}
\end{figure}

\begin{figure*}[!ht]
  \centering
  \includegraphics[width=0.3\linewidth]{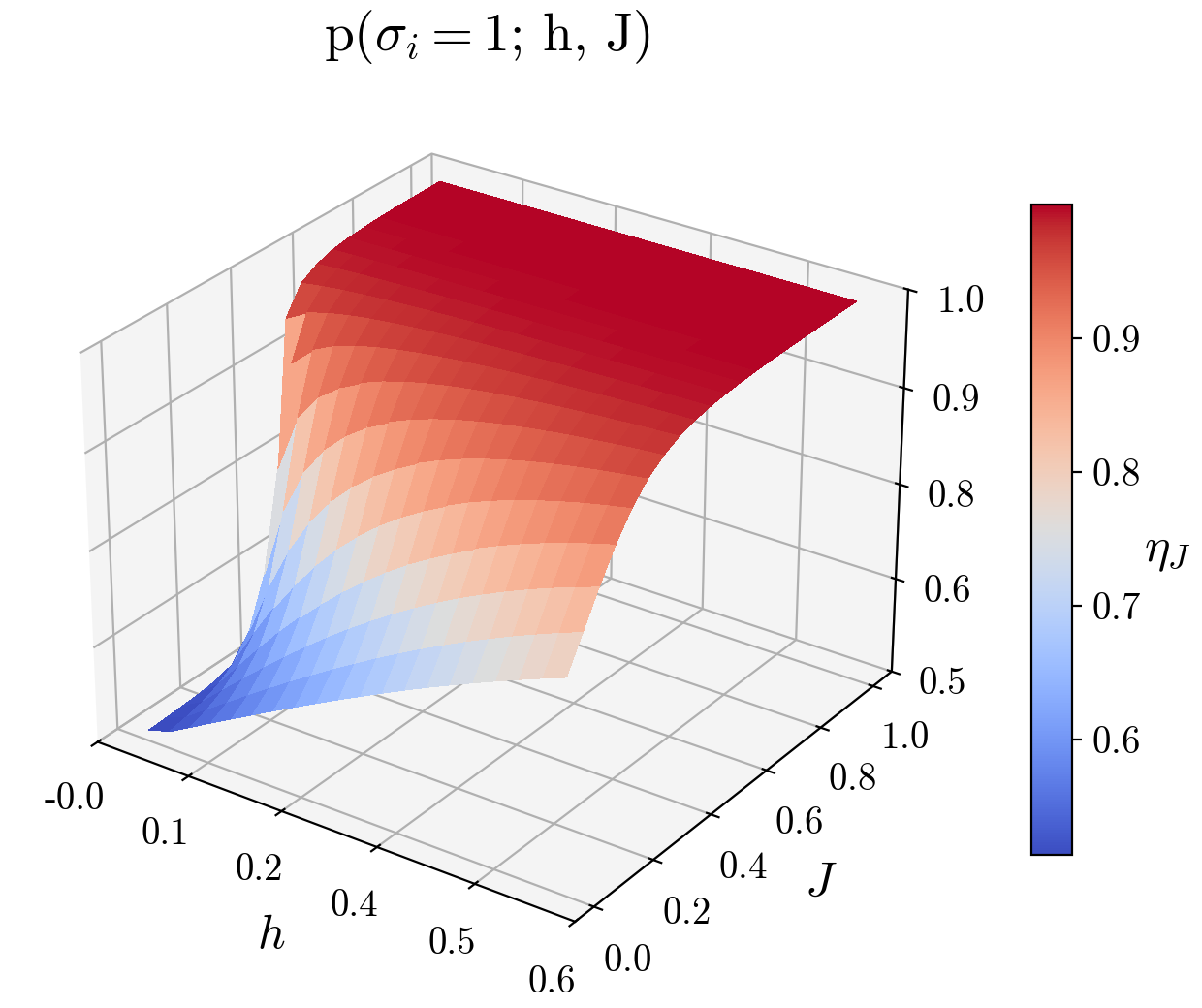}
  \hspace{1em}
  \includegraphics[width=0.3\linewidth]{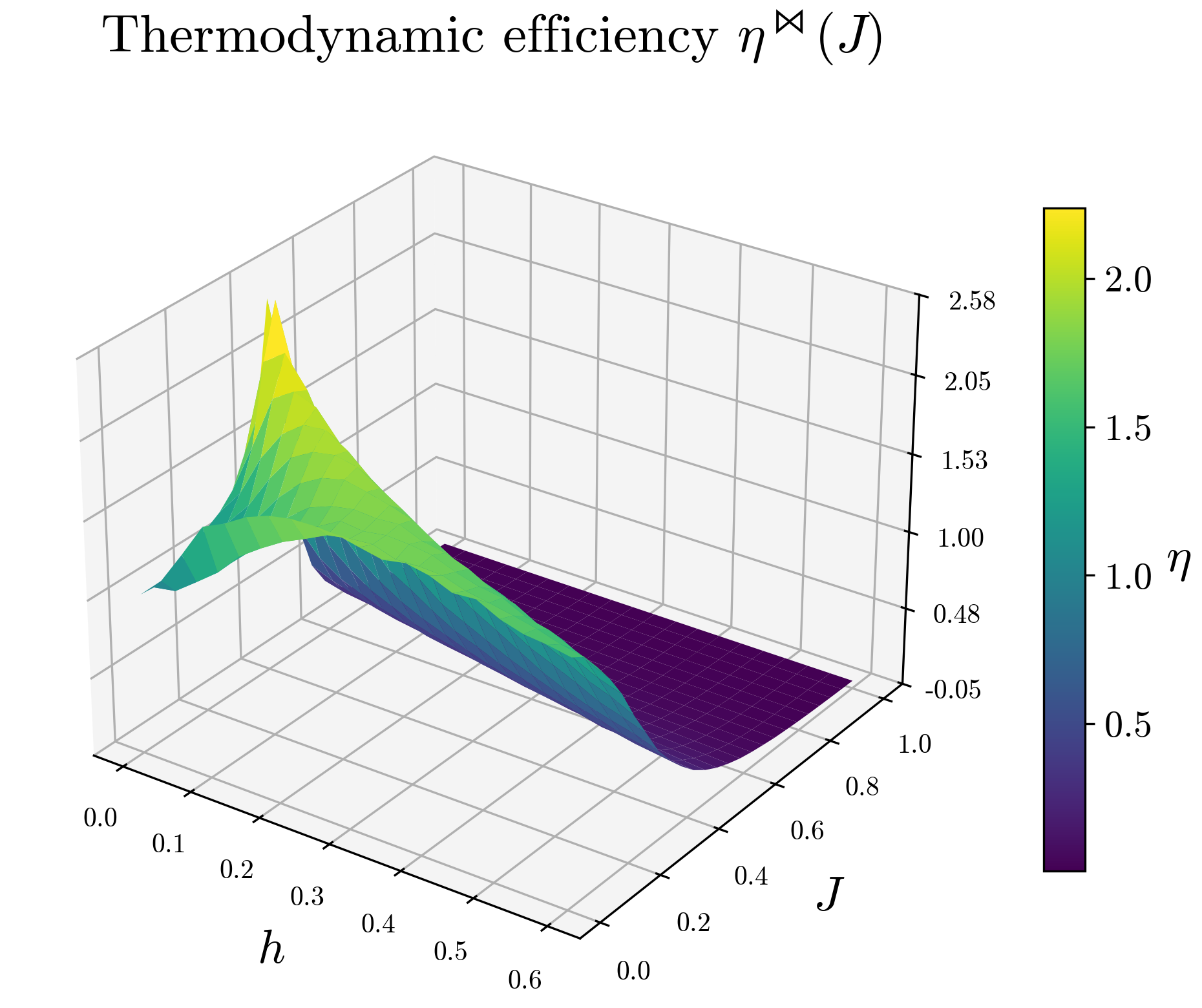}
  \hspace{1em}
  \includegraphics[width=0.3\linewidth]{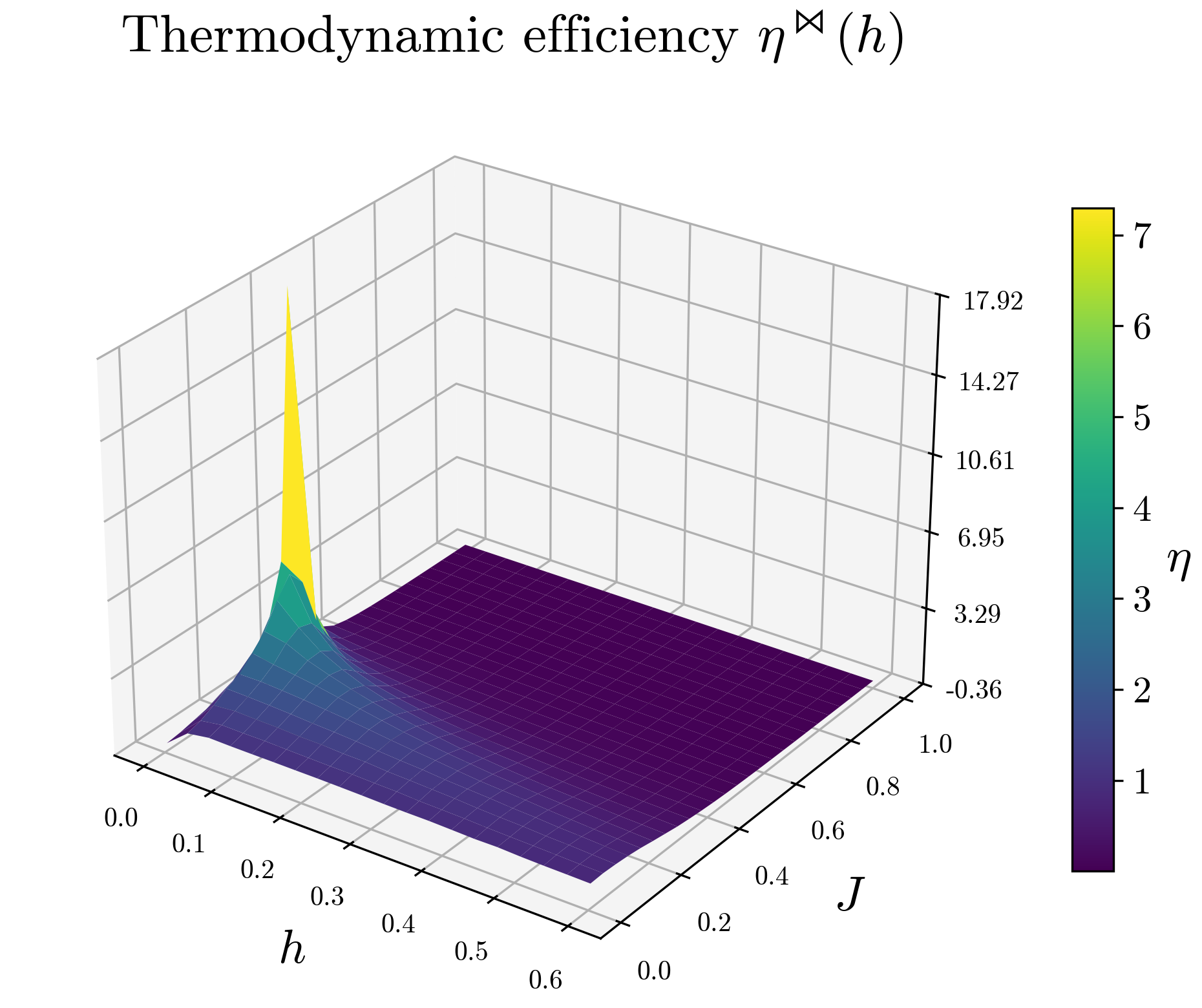}
  \caption{\label{fig:landscape_eta2D} Probability landscape and thermodynamic efficiency $\etaInf$ (inferential form). \textbf{Left}: Probability of positive spin at equilibrium for different combinations of external field strength $h$ and coupling strength $J$. $P(\sigma_i=1)$ closer to 0.5 (blue colour) indicates a disordered state of the system, where spins are equally likely to be up or down; $P(\sigma_i=1)$ closer to 1 (red colour) indicates an ordered state, where spins are predominantly up. \textbf{Middle and Right}: Thermodynamic efficiency $\etaInf$ with respect to $J$ and $h$ (inferential form) for the same parameter combinations. Notice that the edge where the transition from disorder to order is steep corresponds to high thermodynamic efficiency.}
\end{figure*}

Figure \ref{fig:eta_h0} shows simulation results for thermodynamic efficiency with respect to parameter $J$ given external magnetic field $h=0$. When computed using the computational form $\etaCom(J)$ \eqref{eq:eta-ising-derv} peaks slightly beyond the critical point $J_c$ due to the use of a filter required to mitigate noise amplification in differentiating entropy. Moreover, the information-geometric form $\etaGeo(J)$ \eqref{eq:eta-ising-fisher}, which involves Fisher information in both the numerator and denominator, is sensitive to simulation noise, protocol resolution, and system size, making it numerically less stable. As a result, thermodynamic efficiency computed using the information-geometric form $\etaGeo$ often exhibits an inflated peak magnitude relative to the other forms. In comparison, the inferential form $\etaInf(J)$ \eqref{eq:eta-ising-cov} avoids the differentiation operation and is less sensitive to noise, making it a more robust method for numerical simulations. Further details regarding the simulation setup are provided in Appendix~\ref{ap:ising_simulation}.

When we vary both the coupling strength $J$ and the external field $h$, we can evaluate the thermodynamic efficiency across the full parameter space. Figure \ref{fig:landscape_eta2D} shows, for each parameter combination, the equilibrium probability of an up-spin, $P(\sigma_i=1)$ (top) and the thermodynamic efficiency $\etaInf$ with respect to $J$ (middle) and $h$ (right), computed using the inferential form. Note that the probability distribution of spins is not the same as the distribution of microscopic configurations, as multiple configurations may yield the same probability distribution of up and down spins. Nonetheless, the top panel illustrates the regions of order and disorder equilibrium states in the parameter space. The edge with steep transitions from disorder to order equilibrium states aligns with the region of high thermodynamic efficiency. Thermodynamic efficiencies computed using the information-geometric and computational forms yield similar results (see Appendix~\ref{ap:ising_simulation}).

\section{\label{sec:discussion} Conclusion and Discussion}
In this paper, we have considered two different observer-centric interpretations of thermodynamic efficiency by taking the inferential and the information-geometric views, and contrasted them with the previously proposed system-centric, computational view of thermodynamic efficiency. In doing so, we derived two different forms of thermodynamic efficiency while generalising the framework to systems with multiple control parameters, governed by protocols in arbitrary directions.

We first derived the \textit{inferential form}, which is expressed as a normalised, weighted sum of variance and covariances of the macroscopic observables. This formulation captures how changing a specific control parameter $\lambda_j$ influences the joint fluctuations of all the observables $\{X_i\}$, emphasising their inter-dependencies. In this view, thermodynamic efficiency reflects the system's ability to reveal hidden parameters through observable behaviour, thereby connecting the efficiency of self-organization to that of statistical inference. Furthermore, the inferential form provides a more robust and practical method to compute efficiency from simulations.

We also derived the second formulation, namely the \textit{information-geometric form}, and extended thermodynamic efficiency to protocols in arbitrary directions by introducing \textit{directional efficiency}. In this view, directional efficiency is interpreted through the lens of navigation on the statistical manifold following a control protocol trajectory. Given a protocol, directional efficiency evaluates how closely the path direction aligns with the natural gradient of entropy descent, relative to the weighted sum of path energies across all coordinate axes of the manifold. A protocol with high efficiency balances the rate of entropy reduction with the associated total energy. This perspective also connects to parameter space compression, where moving in a stiffer direction leads to a stronger effect on the system's response, corresponding to higher efficiency.

Both interpretations are observer-centric: they focus on the observer's ability to extract information about the system. In contrast, the computational form --- as explored in previous studies \cite{crosato2018Thermodynamics,chen2025Why} --- is system-centric, focusing on the system's ability to perform work to increase predictability in computing its next state. We do not suggest that the system-centric perspective on the thermodynamic efficiency offers an observer-independent measure of the quantity, because estimation of it necessarily requires a particular resolution. However, we argue that thermodynamic efficiency provides an intrinsic utility to the collective system that guides the system's behaviour independent of the observer. The two observer-centric views considered in our work explicitly capture the efficiency with respect to observable quantities, either in terms of accuracy of the statistical inference or in information geometric terms describing measurable perturbations.

Our study assumes the system remains at or near equilibrium, and that the control parameters are varied quasi-statically. This near-equilibrium dynamics leads to exponential family probability distributions of the system's states, which is important for the aforementioned formulations to hold. In addition to the well-known Boltzmann-Gibbs distribution, log-normal distributions are also used in studies of near-equilibrium dynamics, for example, in the context of weak dissipation or multiplicative phenomena \cite{KLEIDON2024173409}. Although the use of exponential family distributions is common in physical, biological and social systems \cite{beck2007Statistics, gustavsson2016Statistical,bialek2012Statistical, nghiem2018Maximumentropy, zanoci2019Ensemble, hunter2007Curved}, other heavier-tailed distributions such as power-law distributions are also used, especially for systems operating far from equilibrium such as financial markets \cite{mantegna1995Scaling, jiang2010Complex}, city population \cite{marsili1998Interacting} and various network systems \cite{barabasi1999Emergence, newman2001Clustering, barabasi2005Origin, muchnik2013Origins}. The extension to non-equilibrium systems is possible, but requires a more careful treatment of the probability distribution and the work done by the system. This will be an important direction for future work, as many real-world systems operate far from equilibrium.

Our findings demonstrate that varying a control parameter affects the efficiency of a self-organizing system through the inter-dependencies of the macroscopic observables. Furthermore, the results bridge the observer-centric and system-centric views of self-organization, and open up new interpretations grounded in statistical inference and information geometry. This unified view deepens our understanding of the fundamental principles governing self-organization and provides a valuable framework for the study of guided self-organization.

\begin{acknowledgments}
Q.C. is supported by the University of Sydney Postgraduate Award (UPA). We wish to acknowledge the support of the Sydney Informatics Hub at the University of Sydney and the National Computing Infrastructure (NCI) Australia for HPC resources that have contributed to the research results reported within this paper.
\end{acknowledgments}

\appendix

\section{Preliminaries and core relations in generalised statistical physics} \label{ap:stat_mech}
The first derivative of log partition function $Z$ with respect to the control parameter $\lambda_j$ gives the expected value of the observable $X_j$ \cite{jaynes1963Information,crooks2007Measuring, niven2010Jaynes}:
\begin{equation} \label{eq:mean-deriv}
    \frac{\partial \log Z(\underline{\lambda})}{\partial \lambda_j} = -\langle X_j \rangle
\end{equation}
while its second derivative gives the covariance of the observables $(X_i, X_j)$ \cite{jaynes1963Information,crooks2007Measuring, niven2010Jaynes}:
\begin{equation} \label{eq:cov-deriv}
    \frac{\partial^2 \log Z}{\partial \lambda_i \partial \lambda_j} = -\frac{\partial \langle X_i \rangle}{\partial \lambda_j} = -\frac{\partial \langle X_j \rangle}{\partial \lambda_i} =\text{Cov}(X_i, X_j)
\end{equation}
It is known that the derivative of entropy can be expressed as follows~\cite[p.~191]{jaynes1963Information}, cf.~\cite[p.~37]{degroot2013Nonequilibrium}:
\begin{eqnarray} \label{eq:numerator1}
  \frac{\partial S(\underline{\lambda})}{\partial \lambda_j} &=& \frac{\partial \log Z}{\partial \lambda_j} + \sum_{i}\lambda_i \frac{\partial \langle X_i \rangle}{\partial \lambda_j} + \sum_{i} \frac{\partial \lambda_i}{\partial \lambda_j}\langle X_i \rangle \nonumber\\
  &=& \sum_{i}\lambda_i \frac{\partial \langle X_i \rangle}{\partial \lambda_j}
\end{eqnarray}
Hence, a combination of \eqref{eq:cov-deriv} and \eqref{eq:numerator1} yields
\begin{equation} \label{eq:numerator_ap}
  \frac{\partial S(\underline{\lambda})}{\partial \lambda_j} = -\sum_{i}\lambda_i \text{Cov}(X_i, X_j)
\end{equation}

Table (\ref{tab:info-vs-thermo}) summarises and compares relevant information-theoretic and statistical physics quantities.

\begin{table*}[ht]
  \centering
  \renewcommand{\arraystretch}{2}%
  \caption{Comparing relevant information-theoretic and statistical physics quantities.}
  \label{tab:info-vs-thermo}
  \begin{ruledtabular}
    \begin{tabular}{l c c c}
      & \textbf{Information Theory} & & \textbf{Statistical Physics} \\[6pt]
      \colrule 
      \shortstack[l]{%
        \rule{0pt}{3.5ex}        
        Microstates, observables
      }
        & $x, X_i(x)$
        & 
        & $x, X_i(x)$ \\[6pt]
      \colrule 
      \shortstack[l]{%
        \rule{0pt}{4ex}
        (generalised) control parameter/ \\
        Lagrange multiplier\footnotemark[1]
      }
        & $\lambda_i$
        &
        & $\beta \hat{\lambda}_i$\\[6pt]
      \colrule 
      \shortstack[l]{\vspace*{1ex} (generalised) energy\footnotemark[2]}
        & 
          \shortstack{
              \rule{0pt}{4ex}%
              Weighted cost:\\[4pt]
              $\displaystyle \sum_{i}\lambda_i\,X_i(x)$
          }
        & \shortstack{
            $\displaystyle \lambda_i = \beta\,\hat\lambda_i$\\[4pt]
            $\longleftrightarrow$
          }
        & 
          \shortstack{
            \rule{0pt}{4ex}%
            Hamiltonian $\mathbb{H}(x)$:\\[4pt]
            $\displaystyle \mathbb{H}(x)=\sum_i \hat\lambda_i\,X_i(x)$
          } 
      \\[6pt]
      \colrule 
      \shortstack[l]{%
        \rule{0pt}{4ex}        
        Equilibrium distribution
      }
        & $p(x) = \frac{1}{Z}e^{-\sum_i \lambda_i\,X_i(x)}$
        & 
        & $p(x) = \frac{1}{Z}e^{-\beta \sum_i \hat\lambda_i\,X_i(x)}$
      \\[6pt]
      \colrule 
      \shortstack[l]{\vspace*{1ex} Entropy}
        & 
          \shortstack{
            \rule{0pt}{4ex}%
            Shannon entropy $S$:\\[4pt]
            $\displaystyle S = -\sum_x p(x)\log p(x)$
          }
        & \shortstack{
            $\displaystyle S = \frac{1}{k_B} \mathbb{S}$\footnotemark[3]\\[4pt]
            $\longleftrightarrow$
          }
        & 
          \shortstack{
            \rule{0pt}{4ex}%
            Thermodynamic entropy $\mathbb{S}$:\\[4pt]
            $\displaystyle \mathbb{S} = -k_B\sum_x p(x)\log p(x)$
          }
      \\[6pt]
      \colrule 
      \shortstack[l]{\vspace*{1ex} Potential}
        & 
          \shortstack{
            \rule{0pt}{4ex}%
            Massieu potential $\Psi$:\\[4pt]
            $\displaystyle \Psi = \log Z = -\sum_{i}\lambda_i \langle X_i \rangle + S$
          }
        & \shortstack{
            $\displaystyle \Psi = -\beta \mathbb{G}$\\[4pt]
            $\longleftrightarrow$
          }
        & 
          \shortstack{
            \rule{0pt}{4ex}%
            Gibbs potential $\mathbb{G}$:\\[4pt]
            $\displaystyle \mathbb{G} = -\tfrac{1}{\beta} \log Z = \langle\mathbb{H}\rangle - T\mathbb{S}$
          }
      \\[6pt]
    \end{tabular}
  \end{ruledtabular}
    \footnotetext[1]{$\lambda_i$ is the generalised control parameter, being the product of the control parameter $\hat{\lambda_i}$ and the corresponding Lagrange multiplier, e.g., the inverse temperature $\beta = 1/(k_B T$).}
  \footnotetext[2]{In information-theoretic studies, the generalised energy is often interpreted as a linear combination of specific cost functions, weighted by their conjugate multipliers~\cite{blahut2003computation,wilson2008boltzmann,salge2015zipf}.}
  \footnotetext[3]{An extra scalar of $\log_2(e)$ is required if $\log$ of base 2 is used in Shannon entropy and the natural $\log$ is used in thermodynamic entropy.}
\end{table*}

\section{Fisher information and its numerical approximation}\label{ap:fisher}
Fisher information measures how sensitive the probability distribution is to changes in the control parameter $\lambda$. Let $p(x;\lambda)$ denote the probability density of the random variable $X$, governed by the parameter $\lambda$. The Fisher information $\mathcal{I}(\lambda)$ of the probability distribution $p(x|\lambda)$ is defined as \cite{cover2005Elements}:
\begin{equation}
    \mathcal{I}(\lambda) = \sum_x p(x|\lambda) \left( \frac{d \log p(x|\lambda)}{d \lambda} \right)^2
\end{equation}
and for multiple control parameters, the Fisher information matrix $\mathcal{I}_{ij}(\underline{\lambda})$ is used \cite{cover2005Elements}:
\begin{equation} \label{eq:fi_multi}
    \mathcal{I}_{ij}(\underline{\lambda}) = \sum_{x} p(x|\underline{\lambda}) \frac{\partial \log p(x|\underline{\lambda})}{\partial \lambda_i}\frac{\partial \log p(x|\underline{\lambda})}{\partial \lambda_j}
\end{equation}

The Fisher information matrix has been shown to be the same as the covariance matrix of observables $X_i$ \cite{brody1995Geometrical, crooks2007Measuring}:
\begin{equation} \label{eq:fisher-covariance}
  \mathcal{I}_{ij}(\underline{\lambda}) = \frac{\partial^2 \log Z}{\partial\lambda_i \partial\lambda_j} =\text{Cov}(X_i, X_j) 
\end{equation}

Numerically, the Fisher information matrix is computed using the covariance matrix of the observables instead of using first principles, as the covariance matrix is computationally more stable.

\section{Simulation of Ising model}\label{ap:ising_simulation}
We consider the canonical Ising model to compare the numerical results of the two different forms of thermodynamic efficiency \cite{chen2025ising2D}. 

The simulations are performed on a toroidal lattice of size $50 \times 50$ using Metropolis algorithm \cite{metropolis1953Equation}. We also explored lattice sizes $25 \times 25$ and $75 \times 75$ for comparison. The coupling strength $J$ is varied from 0 to 1 in increments of 0.05, and the external field strength $h$ ranges from 0 to 0.6 in increments of 0.03. For each value of $J$ and $h$ combination, 500 independent simulations are run, and the average result is used to produce the plot. Each simulation runs for a total of 40.2 million time steps (equivalent to 16080 lattice sweeps for a $50 \times 50$ lattice) to ensure the system reaches equilibrium, with the first 40 million steps discarded as transients. 

To compute thermodynamic efficiency, configuration distributions are sampled from the last 200,000 steps at intervals corresponding to one lattice sweep. The average distribution from these samples is used to compute the Fisher information. Configuration entropy is computed by averaging over 500 simulations, each calculated using the final lattice configuration from the simulation using Kikuchi approximation \cite{kikuchi1951Theory}: 
\begin{equation}
  S = S_4 - 2S_2 + S_1
\end{equation}
where $S_k$ is the entropy of the size-$k$ sublattices.

The lattice is initialised in a fully ordered state. At high coupling strength, single spin-flip algorithms (such as Metropolis or Glauber dynamics) risk trapping the system in a local minimum. Initialising at a fully ordered state allows the system to reach equilibrium faster at high coupling strength, without affecting outcomes at low coupling strength. Since only equilibrium states are analysed and transients are excluded, this method gives the same result as using random initialisation but significantly accelerates convergence.

\begin{figure}[!ht]
  \centering
  \includegraphics[width=1\linewidth]{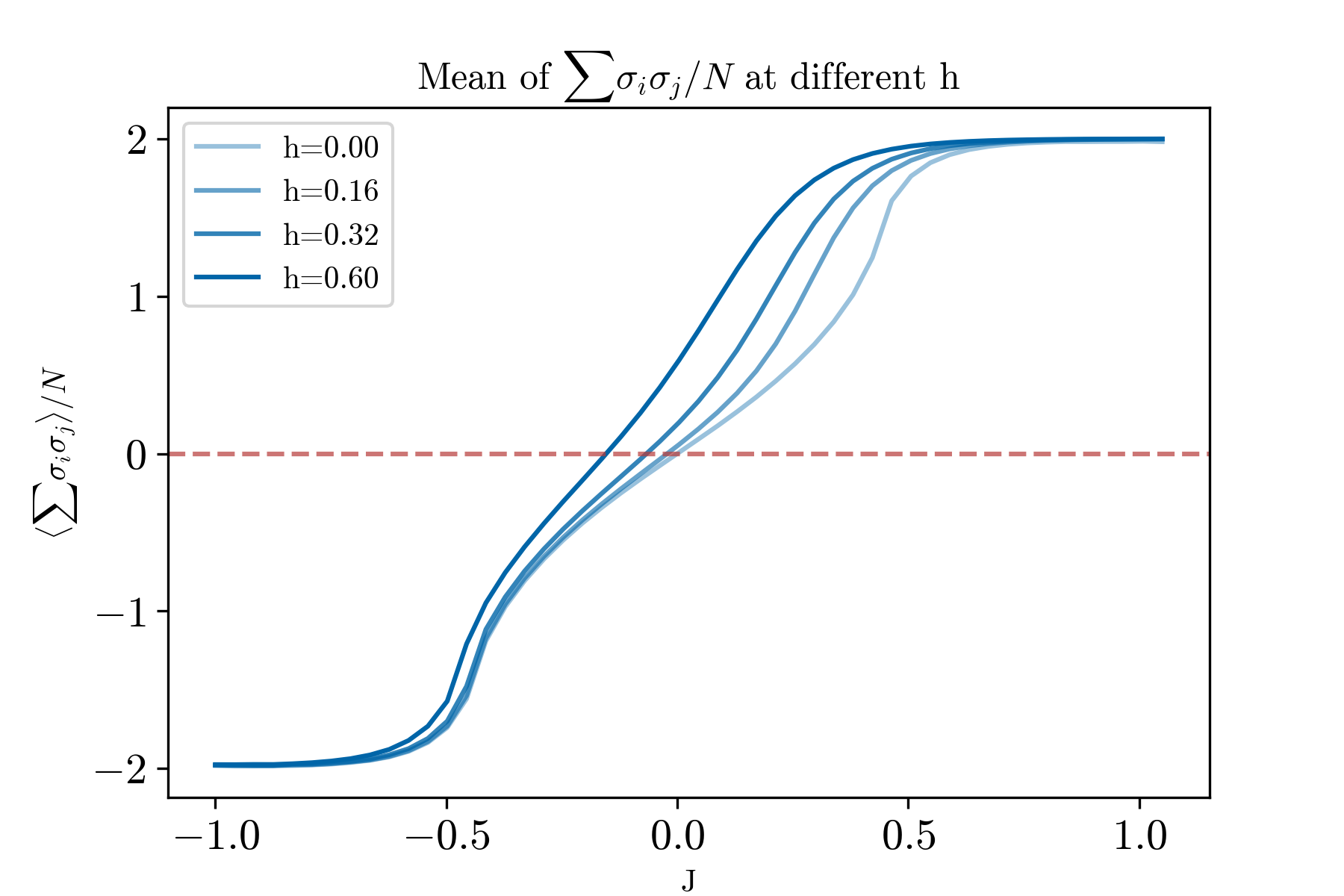}
  \caption{\label{fig:Jstar} 
  Plot of control parameter $J$ against average conjugate observable $\langle \sum \sigma_i \sigma_j\rangle$ (normalised by lattice size), for different values of $h$. The zero-response value $J^*$ is identified as the value of $J$ for which  $\langle \sum \sigma_i \sigma_j\rangle = 0$ (red dotted line). $J^*$ varies with external field $h$.}
\end{figure}

\begin{figure}[!ht]
  \centering
  \includegraphics[width=1\linewidth]{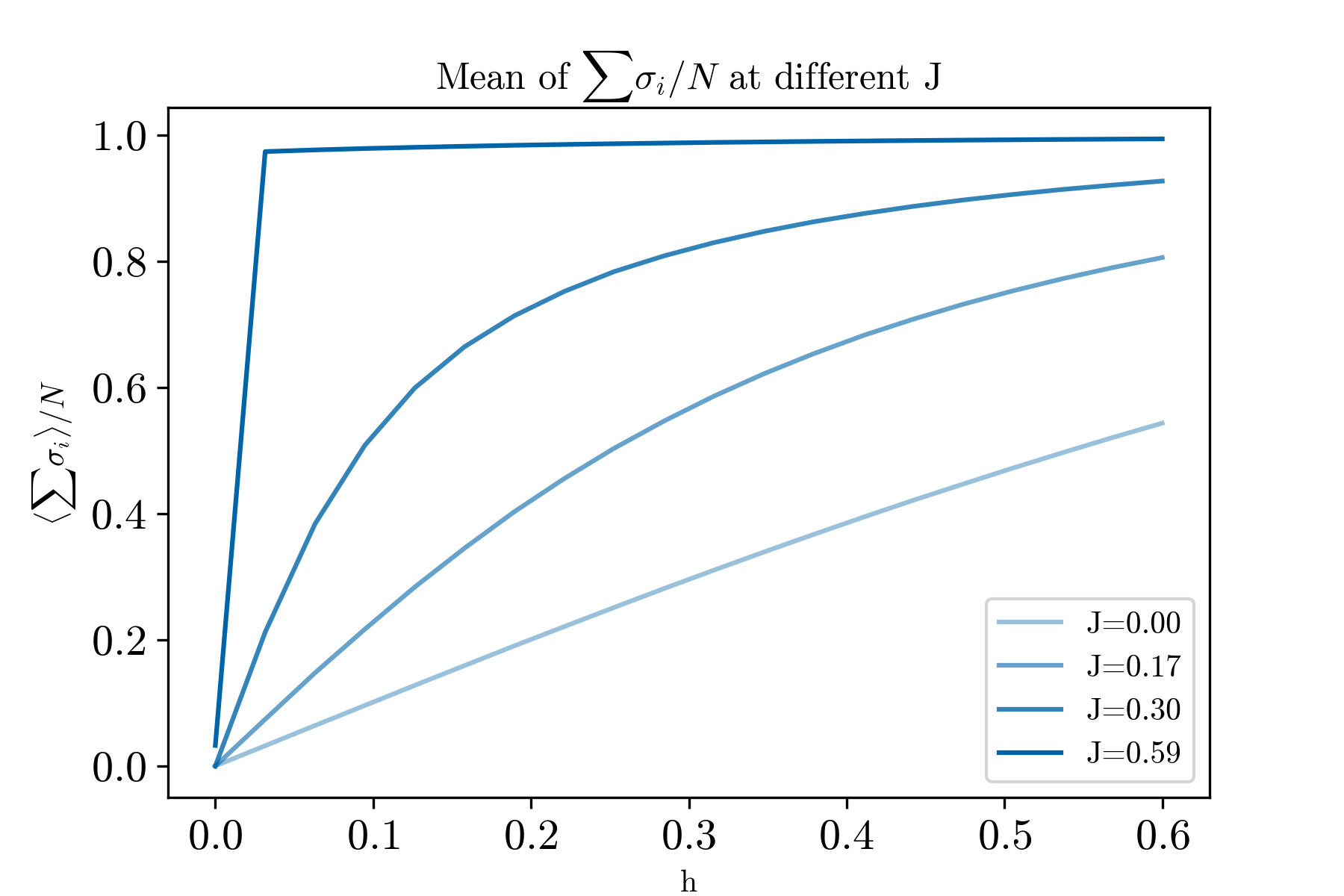}
  \caption{\label{fig:hstar} 
  Plot of control parameter $h$ against average conjugate observable $\langle \sum \sigma_i\rangle$ (normalised by lattice size), for different values of $J$. The zero-response value $h^*$ is identified as the value of $h$ for which  $\langle \sum \sigma_i\rangle = 0$. Due to symmetry in the direction of the external field, $h^*$ is always zero.}
\end{figure}

To compute the integral of Fisher information in computational form $\etaCom$ and information-geometric form $\etaGeo$, one must first identify the zero-response point for each combination of control parameters. The zero-response point for a given control parameter is the value at which the ensemble average of its conjugate observable is zero. For coupling strength $J$, this occurs when the nearest-neighbour correlation is zero (Figure \ref{fig:Jstar}); for the external field $h$, it corresponds to zero average magnetisation (Figure \ref{fig:hstar}).

\begin{figure*}[!ht]
  \centering
  \includegraphics[width=0.4\linewidth]{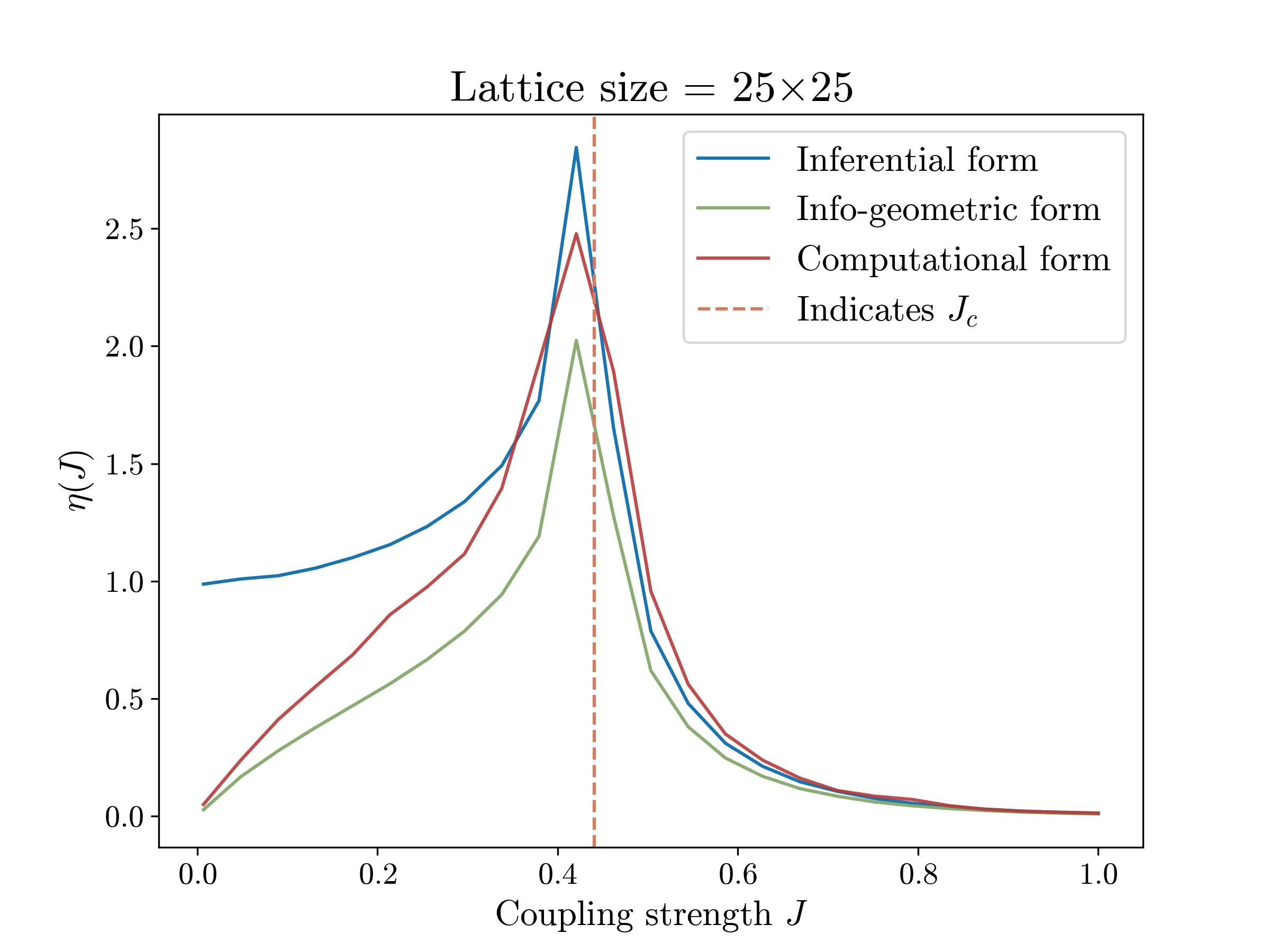}
  \hspace{1em}
  \includegraphics[width=0.4\linewidth]{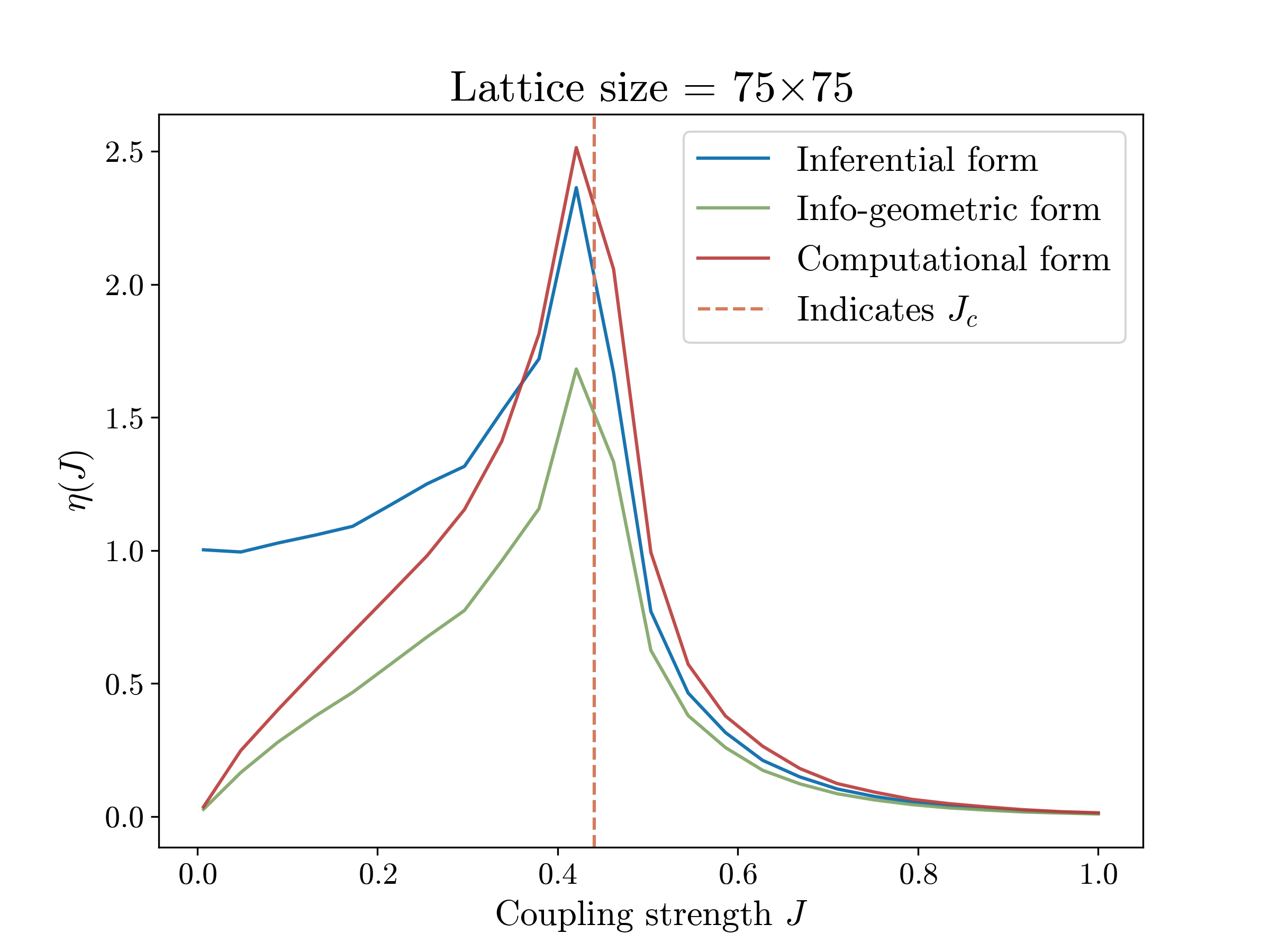}
  \caption{\label{fig:eta_size} Thermodynamic efficiency (computed using three different forms) for the canonical 2D Ising model given zero external magnetic field and various lattice sizes. \textbf{Left}: $25\times25$ lattice. \textbf{Right}: $75\times75$ lattice.}
\end{figure*}

Figure \ref{fig:eta_size} provides a comparison of thermodynamic efficiency across different lattice sizes, simulated with zero external magnetic field. Notice that the information-geometric form, with Fisher information in both the numerator and denominator, is more sensitive to the change of lattice size than the other two forms, giving a much higher value in magnitude. However, all three forms are robust in terms of where the peak occurs (at the proximity of the critical regime).

\begin{figure*}[!ht]
  \centering
  \includegraphics[width=0.4\linewidth]{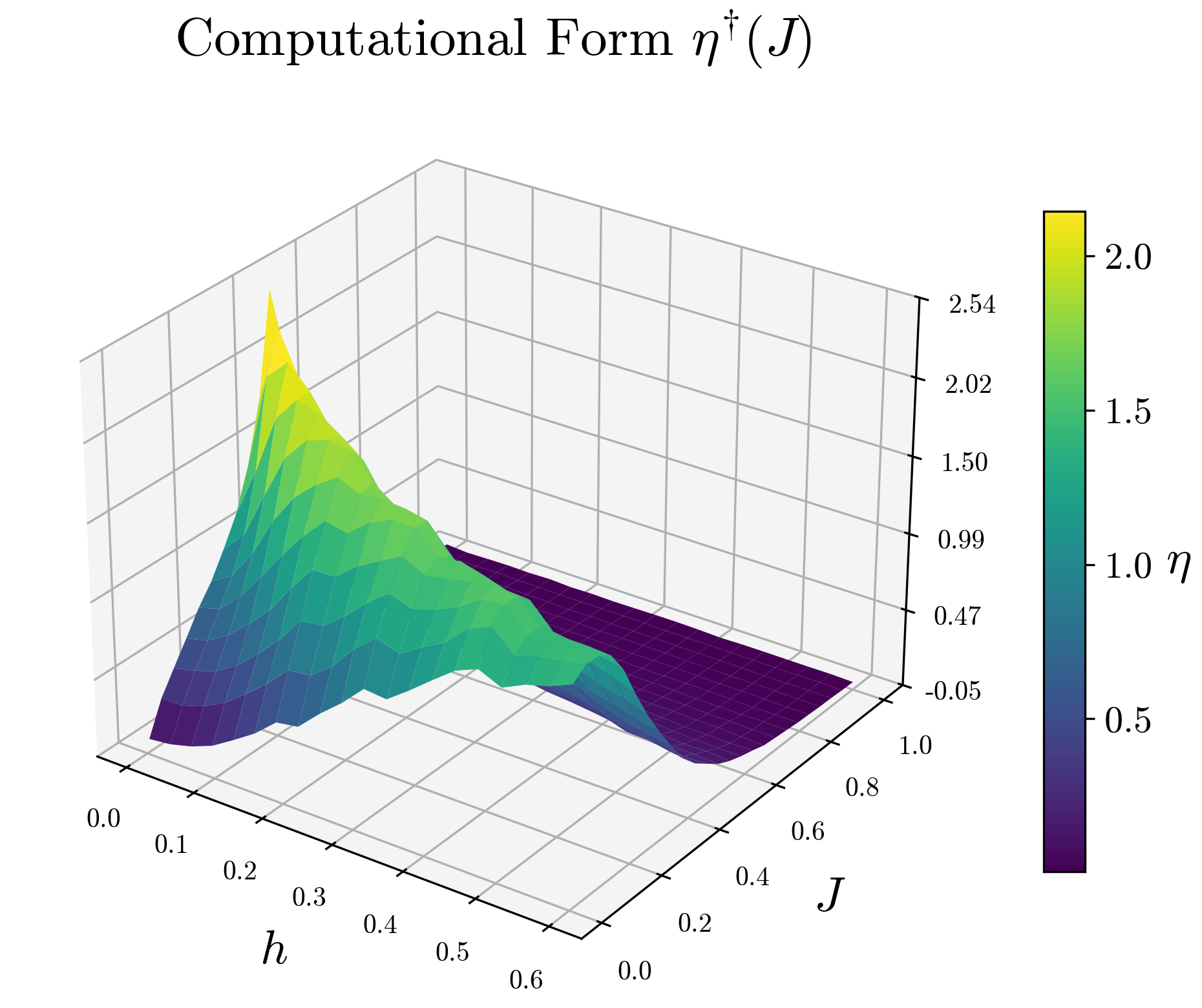}
  \hspace{1em}
  \includegraphics[width=0.4\linewidth]{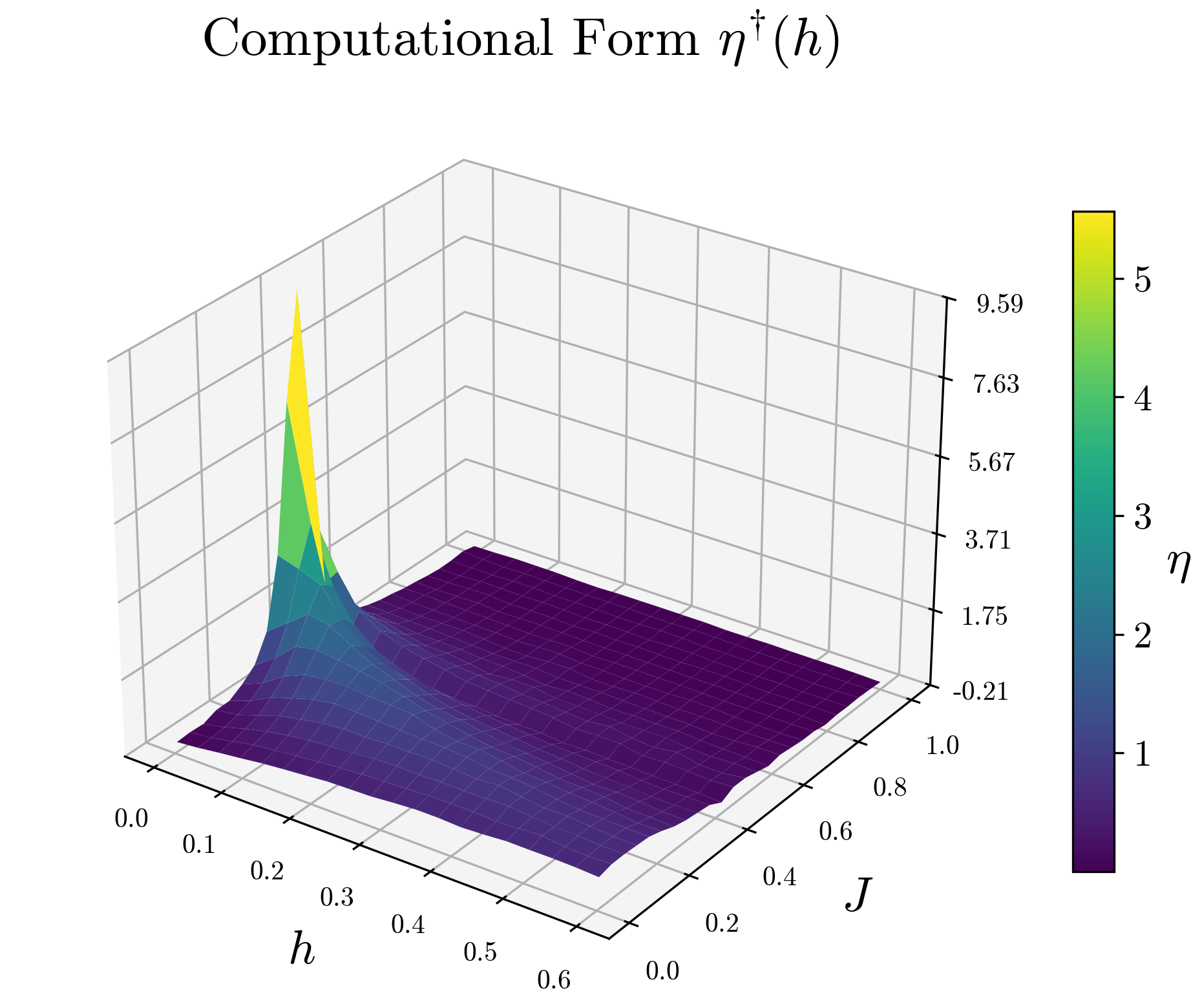}
  \vspace{1em}
  \includegraphics[width=0.4\linewidth]{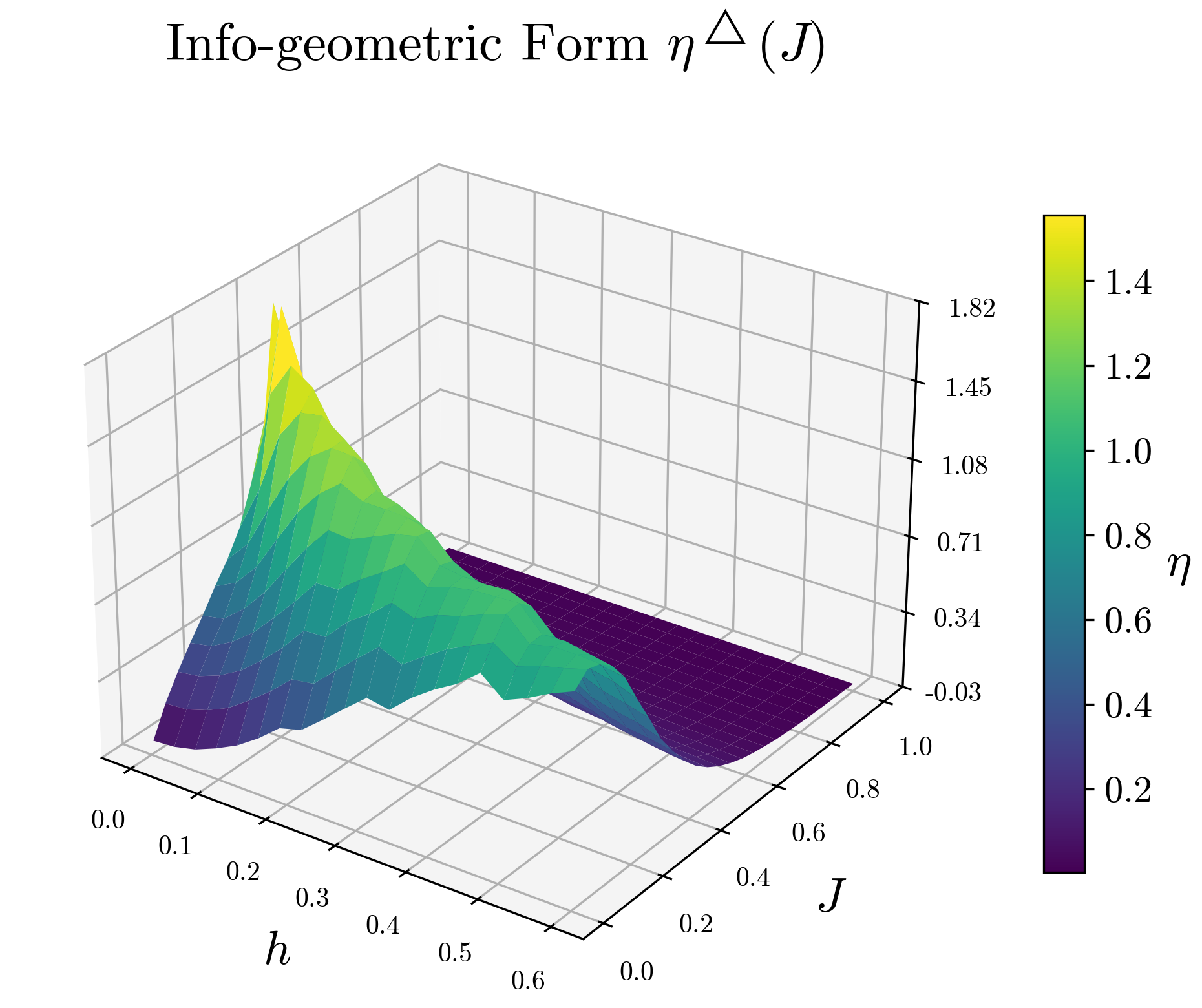}
  \hspace{1em}
  \includegraphics[width=0.4\linewidth]{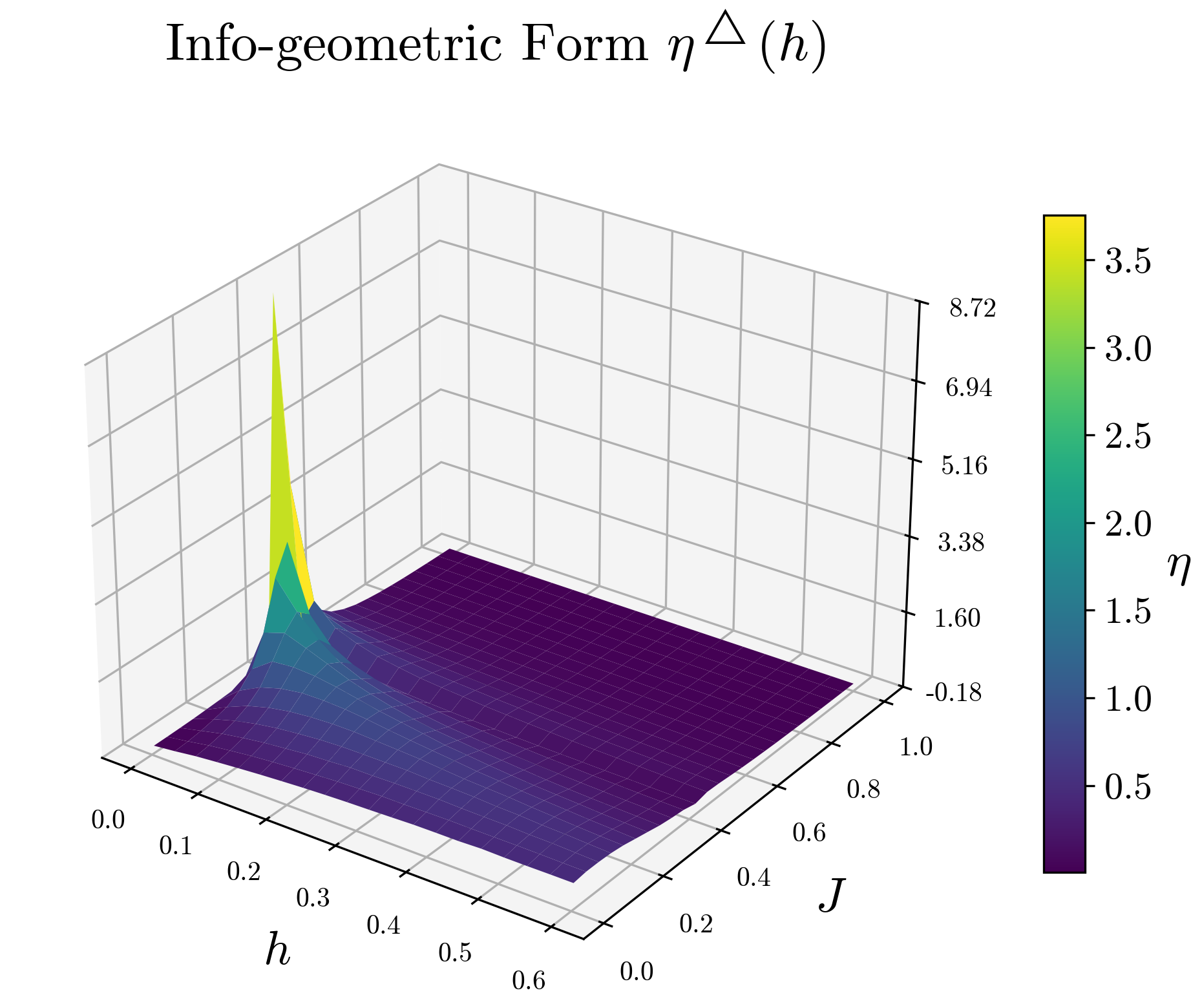}
  \caption{\label{fig:eta2D_allforms} Thermodynamic efficiency of the canonical 2D Ising model computed using computational $\etaCom$ (\textbf{Top}) and info-geometric $\etaGeo$  (\textbf{Bottom}) forms, across a range of coupling strengths $J$, external field strengths $h$. Efficiency is computed separately with respect to each parameter ($J$ and $h$) for a $50\times50$ lattice.}
\end{figure*}

Results for different combinations of external field strengths $h$ and coupling strength $J$ are shown in Figure \ref{fig:eta2D_allforms}. While the methods have different sensitivities to finite-size effects and simulation resolutions, they both indicate the same region of high efficiency, aligning with the sharp transition between ordered and disordered equilibrium states.

\clearpage


\begin{thebibliography}{50}%
\makeatletter
\providecommand \@ifxundefined [1]{%
 \@ifx{#1\undefined}
}%
\providecommand \@ifnum [1]{%
 \ifnum #1\expandafter \@firstoftwo
 \else \expandafter \@secondoftwo
 \fi
}%
\providecommand \@ifx [1]{%
 \ifx #1\expandafter \@firstoftwo
 \else \expandafter \@secondoftwo
 \fi
}%
\providecommand \natexlab [1]{#1}%
\providecommand \enquote  [1]{``#1''}%
\providecommand \bibnamefont  [1]{#1}%
\providecommand \bibfnamefont [1]{#1}%
\providecommand \citenamefont [1]{#1}%
\providecommand \href@noop [0]{\@secondoftwo}%
\providecommand \href [0]{\begingroup \@sanitize@url \@href}%
\providecommand \@href[1]{\@@startlink{#1}\@@href}%
\providecommand \@@href[1]{\endgroup#1\@@endlink}%
\providecommand \@sanitize@url [0]{\catcode `\\12\catcode `\$12\catcode `\&12\catcode `\#12\catcode `\^12\catcode `\_12\catcode `\%12\relax}%
\providecommand \@@startlink[1]{}%
\providecommand \@@endlink[0]{}%
\providecommand \url  [0]{\begingroup\@sanitize@url \@url }%
\providecommand \@url [1]{\endgroup\@href {#1}{\urlprefix }}%
\providecommand \urlprefix  [0]{URL }%
\providecommand \Eprint [0]{\href }%
\providecommand \doibase [0]{https://doi.org/}%
\providecommand \selectlanguage [0]{\@gobble}%
\providecommand \bibinfo  [0]{\@secondoftwo}%
\providecommand \bibfield  [0]{\@secondoftwo}%
\providecommand \translation [1]{[#1]}%
\providecommand \BibitemOpen [0]{}%
\providecommand \bibitemStop [0]{}%
\providecommand \bibitemNoStop [0]{.\EOS\space}%
\providecommand \EOS [0]{\spacefactor3000\relax}%
\providecommand \BibitemShut  [1]{\csname bibitem#1\endcsname}%
\let\auto@bib@innerbib\@empty
\bibitem [{\citenamefont {Ramaswamy}(2010)}]{ramaswamy2010Mechanics}%
  \BibitemOpen
  \bibfield  {author} {\bibinfo {author} {\bibfnamefont {S.}~\bibnamefont {Ramaswamy}},\ }\bibfield  {title} {\bibinfo {title} {The {{Mechanics}} and {{Statistics}} of {{Active Matter}}},\ }\href {https://doi.org/10.1146/annurev-conmatphys-070909-104101} {\bibfield  {journal} {\bibinfo  {journal} {Annual Review of Condensed Matter Physics}\ }\textbf {\bibinfo {volume} {1}},\ \bibinfo {pages} {323} (\bibinfo {year} {2010})}\BibitemShut {NoStop}%
\bibitem [{\citenamefont {Marchetti}\ \emph {et~al.}(2013)\citenamefont {Marchetti}, \citenamefont {Joanny}, \citenamefont {Ramaswamy}, \citenamefont {Liverpool}, \citenamefont {Prost}, \citenamefont {Rao},\ and\ \citenamefont {Simha}}]{marchetti2013Hydrodynamics}%
  \BibitemOpen
  \bibfield  {author} {\bibinfo {author} {\bibfnamefont {M.~C.}\ \bibnamefont {Marchetti}}, \bibinfo {author} {\bibfnamefont {J.~F.}\ \bibnamefont {Joanny}}, \bibinfo {author} {\bibfnamefont {S.}~\bibnamefont {Ramaswamy}}, \bibinfo {author} {\bibfnamefont {T.~B.}\ \bibnamefont {Liverpool}}, \bibinfo {author} {\bibfnamefont {J.}~\bibnamefont {Prost}}, \bibinfo {author} {\bibfnamefont {M.}~\bibnamefont {Rao}},\ and\ \bibinfo {author} {\bibfnamefont {R.~A.}\ \bibnamefont {Simha}},\ }\bibfield  {title} {\bibinfo {title} {Hydrodynamics of soft active matter},\ }\href {https://doi.org/10.1103/RevModPhys.85.1143} {\bibfield  {journal} {\bibinfo  {journal} {Reviews of Modern Physics}\ }\textbf {\bibinfo {volume} {85}},\ \bibinfo {pages} {1143} (\bibinfo {year} {2013})}\BibitemShut {NoStop}%
\bibitem [{\citenamefont {Vicsek}\ and\ \citenamefont {Zafeiris}(2012)}]{vicsek2012Collective}%
  \BibitemOpen
  \bibfield  {author} {\bibinfo {author} {\bibfnamefont {T.}~\bibnamefont {Vicsek}}\ and\ \bibinfo {author} {\bibfnamefont {A.}~\bibnamefont {Zafeiris}},\ }\bibfield  {title} {\bibinfo {title} {Collective motion},\ }\href {https://doi.org/10.1016/j.physrep.2012.03.004} {\bibfield  {journal} {\bibinfo  {journal} {Physics Reports}\ }\textbf {\bibinfo {volume} {517}},\ \bibinfo {pages} {71} (\bibinfo {year} {2012})},\ \Eprint {https://arxiv.org/abs/1010.5017} {arXiv:1010.5017} \BibitemShut {NoStop}%
\bibitem [{\citenamefont {Crosato}\ \emph {et~al.}(2019)\citenamefont {Crosato}, \citenamefont {Prokopenko},\ and\ \citenamefont {Spinney}}]{crosato2019irreversibility}%
  \BibitemOpen
  \bibfield  {author} {\bibinfo {author} {\bibfnamefont {E.}~\bibnamefont {Crosato}}, \bibinfo {author} {\bibfnamefont {M.}~\bibnamefont {Prokopenko}},\ and\ \bibinfo {author} {\bibfnamefont {R.~E.}\ \bibnamefont {Spinney}},\ }\bibfield  {title} {\bibinfo {title} {Irreversibility and emergent structure in active matter},\ }\href {https://doi.org/10.1103/PhysRevE.100.042613} {\bibfield  {journal} {\bibinfo  {journal} {Physical Review E}\ }\textbf {\bibinfo {volume} {100}},\ \bibinfo {pages} {042613} (\bibinfo {year} {2019})}\BibitemShut {NoStop}%
\bibitem [{\citenamefont {Crosato}\ \emph {et~al.}(2018{\natexlab{a}})\citenamefont {Crosato}, \citenamefont {Nigmatullin},\ and\ \citenamefont {Prokopenko}}]{crosato2018Critical}%
  \BibitemOpen
  \bibfield  {author} {\bibinfo {author} {\bibfnamefont {E.}~\bibnamefont {Crosato}}, \bibinfo {author} {\bibfnamefont {R.}~\bibnamefont {Nigmatullin}},\ and\ \bibinfo {author} {\bibfnamefont {M.}~\bibnamefont {Prokopenko}},\ }\bibfield  {title} {\bibinfo {title} {On critical dynamics and thermodynamic efficiency of urban transformations},\ }\bibfield  {journal} {\bibinfo  {journal} {Royal Society Open Science}\ }\textbf {\bibinfo {volume} {5}},\ \href {https://doi.org/10.1098/rsos.180863} {10.1098/rsos.180863} (\bibinfo {year} {2018}{\natexlab{a}})\BibitemShut {NoStop}%
\bibitem [{\citenamefont {Crosato}\ \emph {et~al.}(2018{\natexlab{b}})\citenamefont {Crosato}, \citenamefont {Spinney}, \citenamefont {Nigmatullin}, \citenamefont {Lizier},\ and\ \citenamefont {Prokopenko}}]{crosato2018Thermodynamics}%
  \BibitemOpen
  \bibfield  {author} {\bibinfo {author} {\bibfnamefont {E.}~\bibnamefont {Crosato}}, \bibinfo {author} {\bibfnamefont {R.~E.}\ \bibnamefont {Spinney}}, \bibinfo {author} {\bibfnamefont {R.}~\bibnamefont {Nigmatullin}}, \bibinfo {author} {\bibfnamefont {J.~T.}\ \bibnamefont {Lizier}},\ and\ \bibinfo {author} {\bibfnamefont {M.}~\bibnamefont {Prokopenko}},\ }\bibfield  {title} {\bibinfo {title} {Thermodynamics and computation during collective motion near criticality},\ }\href {https://doi.org/10.1103/PhysRevE.97.012120} {\bibfield  {journal} {\bibinfo  {journal} {Physical Review E}\ }\textbf {\bibinfo {volume} {97}},\ \bibinfo {pages} {1} (\bibinfo {year} {2018}{\natexlab{b}})}\BibitemShut {NoStop}%
\bibitem [{\citenamefont {Harding}\ \emph {et~al.}(2018)\citenamefont {Harding}, \citenamefont {Nigmatullin},\ and\ \citenamefont {Prokopenko}}]{harding2018Thermodynamic}%
  \BibitemOpen
  \bibfield  {author} {\bibinfo {author} {\bibfnamefont {N.}~\bibnamefont {Harding}}, \bibinfo {author} {\bibfnamefont {R.}~\bibnamefont {Nigmatullin}},\ and\ \bibinfo {author} {\bibfnamefont {M.}~\bibnamefont {Prokopenko}},\ }\bibfield  {title} {\bibinfo {title} {Thermodynamic efficiency of contagions: {{A}} statistical mechanical analysis of the {{SIS}} epidemic model},\ }\href {https://doi.org/10.1098/rsfs.2018.0036} {\bibfield  {journal} {\bibinfo  {journal} {Interface Focus}\ }\textbf {\bibinfo {volume} {8}},\ \bibinfo {pages} {20180036} (\bibinfo {year} {2018})}\BibitemShut {NoStop}%
\bibitem [{\citenamefont {Nigmatullin}\ and\ \citenamefont {Prokopenko}(2021)}]{nigmatullin2021Thermodynamic}%
  \BibitemOpen
  \bibfield  {author} {\bibinfo {author} {\bibfnamefont {R.}~\bibnamefont {Nigmatullin}}\ and\ \bibinfo {author} {\bibfnamefont {M.}~\bibnamefont {Prokopenko}},\ }\bibfield  {title} {\bibinfo {title} {Thermodynamic efficiency of interactions in self-organizing systems},\ }\href {https://doi.org/10.3390/e23060757} {\bibfield  {journal} {\bibinfo  {journal} {Entropy}\ }\textbf {\bibinfo {volume} {23}},\ \bibinfo {pages} {757} (\bibinfo {year} {2021})}\BibitemShut {NoStop}%
\bibitem [{\citenamefont {Chen}\ and\ \citenamefont {Prokopenko}(2025)}]{chen2025Why}%
  \BibitemOpen
  \bibfield  {author} {\bibinfo {author} {\bibfnamefont {Q.}~\bibnamefont {Chen}}\ and\ \bibinfo {author} {\bibfnamefont {M.}~\bibnamefont {Prokopenko}},\ }\bibfield  {title} {\bibinfo {title} {Why collective behaviours self-organize to criticality: A primer on information-theoretic and thermodynamic utility measures},\ }\href {https://doi.org/10.1098/rsos.241655} {\bibfield  {journal} {\bibinfo  {journal} {Royal Society Open Science}\ }\textbf {\bibinfo {volume} {12}},\ \bibinfo {pages} {241655} (\bibinfo {year} {2025})}\BibitemShut {NoStop}%
\bibitem [{\citenamefont {Jaynes}(1957)}]{jaynes1957Information}%
  \BibitemOpen
  \bibfield  {author} {\bibinfo {author} {\bibfnamefont {E.~T.}\ \bibnamefont {Jaynes}},\ }\bibfield  {title} {\bibinfo {title} {Information {{Theory}} and {{Statistical Mechanics}}},\ }\href {https://doi.org/10.1103/PhysRev.106.620} {\bibfield  {journal} {\bibinfo  {journal} {Physical Review}\ }\textbf {\bibinfo {volume} {106}},\ \bibinfo {pages} {620} (\bibinfo {year} {1957})}\BibitemShut {NoStop}%
\bibitem [{\citenamefont {Crooks}(2007)}]{crooks2007Measuring}%
  \BibitemOpen
  \bibfield  {author} {\bibinfo {author} {\bibfnamefont {G.~E.}\ \bibnamefont {Crooks}},\ }\bibfield  {title} {\bibinfo {title} {Measuring {{Thermodynamic Length}}},\ }\href {https://doi.org/10.1103/PhysRevLett.99.100602} {\bibfield  {journal} {\bibinfo  {journal} {Physical Review Letters}\ }\textbf {\bibinfo {volume} {99}},\ \bibinfo {pages} {100602} (\bibinfo {year} {2007})}\BibitemShut {NoStop}%
\bibitem [{\citenamefont {Shannon}(1948)}]{shannon1948Mathematical}%
  \BibitemOpen
  \bibfield  {author} {\bibinfo {author} {\bibfnamefont {C.~E.}\ \bibnamefont {Shannon}},\ }\bibfield  {title} {\bibinfo {title} {A {{Mathematical Theory}} of {{Communication}}},\ }\href {https://doi.org/10.1002/j.1538-7305.1948.tb00917.x} {\bibfield  {journal} {\bibinfo  {journal} {Bell System Technical Journal}\ }\textbf {\bibinfo {volume} {27}},\ \bibinfo {pages} {623} (\bibinfo {year} {1948})}\BibitemShut {NoStop}%
\bibitem [{\citenamefont {Landau}\ and\ \citenamefont {Lif{\v s}ic}(2011)}]{landau2011Statistical}%
  \BibitemOpen
  \bibfield  {author} {\bibinfo {author} {\bibfnamefont {L.~D.}\ \bibnamefont {Landau}}\ and\ \bibinfo {author} {\bibfnamefont {E.~M.}\ \bibnamefont {Lif{\v s}ic}},\ }\href@noop {} {\emph {\bibinfo {title} {Statistical Physics}}},\ \bibinfo {edition} {3rd}\ ed.,\ \bibinfo {series} {Course of Theoretical Physics}, Vol.~\bibinfo {volume} {5}\ (\bibinfo  {publisher} {Elsevier Butterworth Heinemann},\ \bibinfo {address} {Amsterdam Heidelberg},\ \bibinfo {year} {2011})\BibitemShut {NoStop}%
\bibitem [{\citenamefont {Loomis}\ and\ \citenamefont {Sternberg}(1990)}]{loomis1990Advanced}%
  \BibitemOpen
  \bibfield  {author} {\bibinfo {author} {\bibfnamefont {L.~H.}\ \bibnamefont {Loomis}}\ and\ \bibinfo {author} {\bibfnamefont {S.}~\bibnamefont {Sternberg}},\ }\href@noop {} {\emph {\bibinfo {title} {Advanced Calculus}}},\ \bibinfo {edition} {rev. ed}\ ed.\ (\bibinfo  {publisher} {{Jones and Bartlett}},\ \bibinfo {address} {Boston},\ \bibinfo {year} {1990})\BibitemShut {NoStop}%
\bibitem [{\citenamefont {Jaynes}(1963)}]{jaynes1963Information}%
  \BibitemOpen
  \bibfield  {author} {\bibinfo {author} {\bibfnamefont {E.}~\bibnamefont {Jaynes}},\ }\bibfield  {title} {\bibinfo {title} {Information theory and statistical mechanics},\ }in\ \href@noop {} {\emph {\bibinfo {booktitle} {Lectures in Theoretical Physics, {{Vol}}. 3: {{Statistical Physics}}}}},\ \bibinfo {series and number} {Brandeis {{University Summer Institute}}},\ \bibinfo {editor} {edited by\ \bibinfo {editor} {\bibfnamefont {K.}~\bibnamefont {Ford}}}\ (\bibinfo  {publisher} {W. A. Benjamin, Inc.},\ \bibinfo {address} {New York},\ \bibinfo {year} {1963})\ pp.\ \bibinfo {pages} {181--218}\BibitemShut {NoStop}%
\bibitem [{\citenamefont {Fisher}(1922)}]{fisher1922Mathematical}%
  \BibitemOpen
  \bibfield  {author} {\bibinfo {author} {\bibfnamefont {R.~A.}\ \bibnamefont {Fisher}},\ }\bibfield  {title} {\bibinfo {title} {On the mathematical foundations of theoretical statistics},\ }\href {https://doi.org/10.1098/rsta.1922.0009} {\bibfield  {journal} {\bibinfo  {journal} {Philosophical Transactions of the Royal Society of London. Series A, Containing Papers of a Mathematical or Physical Character}\ }\textbf {\bibinfo {volume} {222}},\ \bibinfo {pages} {309} (\bibinfo {year} {1922})}\BibitemShut {NoStop}%
\bibitem [{\citenamefont {Cover}\ and\ \citenamefont {Thomas}(2005)}]{cover2005Elements}%
  \BibitemOpen
  \bibfield  {author} {\bibinfo {author} {\bibfnamefont {T.~M.}\ \bibnamefont {Cover}}\ and\ \bibinfo {author} {\bibfnamefont {J.~A.}\ \bibnamefont {Thomas}},\ }\href@noop {} {\emph {\bibinfo {title} {Elements of Information Theory}}},\ \bibinfo {edition} {2nd}\ ed.\ (\bibinfo  {publisher} {John Wiley \& Sons, Inc.},\ \bibinfo {address} {Hoboken, New Jersey},\ \bibinfo {year} {2005})\BibitemShut {NoStop}%
\bibitem [{\citenamefont {Prokopenko}\ \emph {et~al.}(2011)\citenamefont {Prokopenko}, \citenamefont {Lizier}, \citenamefont {Obst},\ and\ \citenamefont {Wang}}]{prokopenko2011Relating}%
  \BibitemOpen
  \bibfield  {author} {\bibinfo {author} {\bibfnamefont {M.}~\bibnamefont {Prokopenko}}, \bibinfo {author} {\bibfnamefont {J.~T.}\ \bibnamefont {Lizier}}, \bibinfo {author} {\bibfnamefont {O.}~\bibnamefont {Obst}},\ and\ \bibinfo {author} {\bibfnamefont {X.~R.}\ \bibnamefont {Wang}},\ }\bibfield  {title} {\bibinfo {title} {Relating {{Fisher}} information to order parameters},\ }\href {https://doi.org/10.1103/PhysRevE.84.041116} {\bibfield  {journal} {\bibinfo  {journal} {Physical Review E - Statistical, Nonlinear, and Soft Matter Physics}\ }\textbf {\bibinfo {volume} {84}},\ \bibinfo {pages} {041116} (\bibinfo {year} {2011})}\BibitemShut {NoStop}%
\bibitem [{\citenamefont {Amari}\ and\ \citenamefont {Nagaoka}(2007)}]{amari2007Methods}%
  \BibitemOpen
  \bibfield  {author} {\bibinfo {author} {\bibfnamefont {S.-I.}\ \bibnamefont {Amari}}\ and\ \bibinfo {author} {\bibfnamefont {H.}~\bibnamefont {Nagaoka}},\ }\href@noop {} {\emph {\bibinfo {title} {Methods of Information Geometry}}},\ \bibinfo {series} {Translations of Mathematical Monographs}\ No.\ \bibinfo {number} {191}\ (\bibinfo  {publisher} {American Mathematical Society},\ \bibinfo {address} {Providence, Rhode Island},\ \bibinfo {year} {2007})\BibitemShut {NoStop}%
\bibitem [{\citenamefont {Amari}(2016)}]{amari2016Information}%
  \BibitemOpen
  \bibfield  {author} {\bibinfo {author} {\bibfnamefont {S.-I.}\ \bibnamefont {Amari}},\ }\href {https://doi.org/10.1007/978-4-431-55978-8} {\emph {\bibinfo {title} {Information {{Geometry}} and {{Its Applications}}}}},\ \bibinfo {series} {Applied {{Mathematical Sciences}}}, Vol.\ \bibinfo {volume} {194}\ (\bibinfo  {publisher} {Springer Japan},\ \bibinfo {address} {Tokyo},\ \bibinfo {year} {2016})\BibitemShut {NoStop}%
\bibitem [{\citenamefont {Ay}\ \emph {et~al.}(2017)\citenamefont {Ay}, \citenamefont {Jost}, \citenamefont {L{\^e}},\ and\ \citenamefont {Schwachh{\"o}fer}}]{ay2017Information}%
  \BibitemOpen
  \bibfield  {author} {\bibinfo {author} {\bibfnamefont {N.}~\bibnamefont {Ay}}, \bibinfo {author} {\bibfnamefont {J.}~\bibnamefont {Jost}}, \bibinfo {author} {\bibfnamefont {H.~V.}\ \bibnamefont {L{\^e}}},\ and\ \bibinfo {author} {\bibfnamefont {L.}~\bibnamefont {Schwachh{\"o}fer}},\ }\href {https://doi.org/10.1007/978-3-319-56478-4} {\emph {\bibinfo {title} {Information Geometry}}},\ \bibinfo {series} {Ergebnisse der Mathematik und ihrer Grenzgebiete. 3. Folge / A Series of Modern Surveys in Mathematics}, Vol.~\bibinfo {volume} {64}\ (\bibinfo  {publisher} {Springer International Publishing},\ \bibinfo {year} {2017})\BibitemShut {NoStop}%
\bibitem [{\citenamefont {Amari}(2021)}]{amari2021Information}%
  \BibitemOpen
  \bibfield  {author} {\bibinfo {author} {\bibfnamefont {S.-I.}\ \bibnamefont {Amari}},\ }\bibfield  {title} {\bibinfo {title} {Information {{Geometry}}},\ }\href {https://doi.org/10.1111/insr.12464} {\bibfield  {journal} {\bibinfo  {journal} {International Statistical Review}\ }\textbf {\bibinfo {volume} {89}},\ \bibinfo {pages} {250} (\bibinfo {year} {2021})}\BibitemShut {NoStop}%
\bibitem [{\citenamefont {Amari}(1998)}]{amari1998natural}%
  \BibitemOpen
  \bibfield  {author} {\bibinfo {author} {\bibfnamefont {S.-I.}\ \bibnamefont {Amari}},\ }\bibfield  {title} {\bibinfo {title} {Natural gradient works efficiently in learning},\ }\href@noop {} {\bibfield  {journal} {\bibinfo  {journal} {Neural computation}\ }\textbf {\bibinfo {volume} {10}},\ \bibinfo {pages} {251} (\bibinfo {year} {1998})}\BibitemShut {NoStop}%
\bibitem [{\citenamefont {Nielsen}(2020)}]{nielsen2020Elementary}%
  \BibitemOpen
  \bibfield  {author} {\bibinfo {author} {\bibfnamefont {F.}~\bibnamefont {Nielsen}},\ }\bibfield  {title} {\bibinfo {title} {An elementary introduction to information geometry},\ }\href {https://doi.org/10.3390/e22101100} {\bibfield  {journal} {\bibinfo  {journal} {Entropy}\ }\textbf {\bibinfo {volume} {22}},\ \bibinfo {pages} {1100} (\bibinfo {year} {2020})},\ \Eprint {https://arxiv.org/abs/1808.08271} {1808.08271} \BibitemShut {NoStop}%
\bibitem [{\citenamefont {Jost}(2017)}]{jost2017Riemannian}%
  \BibitemOpen
  \bibfield  {author} {\bibinfo {author} {\bibfnamefont {J.}~\bibnamefont {Jost}},\ }\href {https://doi.org/10.1007/978-3-319-61860-9} {\emph {\bibinfo {title} {Riemannian Geometry and Geometric Analysis}}},\ Universitext\ (\bibinfo  {publisher} {Springer International Publishing},\ \bibinfo {year} {2017})\BibitemShut {NoStop}%
\bibitem [{\citenamefont {Machta}\ \emph {et~al.}(2013)\citenamefont {Machta}, \citenamefont {Chachra}, \citenamefont {Transtrum},\ and\ \citenamefont {Sethna}}]{machta2013Parameter}%
  \BibitemOpen
  \bibfield  {author} {\bibinfo {author} {\bibfnamefont {B.~B.}\ \bibnamefont {Machta}}, \bibinfo {author} {\bibfnamefont {R.}~\bibnamefont {Chachra}}, \bibinfo {author} {\bibfnamefont {M.~K.}\ \bibnamefont {Transtrum}},\ and\ \bibinfo {author} {\bibfnamefont {J.~P.}\ \bibnamefont {Sethna}},\ }\bibfield  {title} {\bibinfo {title} {Parameter {{Space Compression Underlies Emergent Theories}} and {{Predictive Models}}},\ }\href {https://doi.org/10.1126/science.1238723} {\bibfield  {journal} {\bibinfo  {journal} {Science}\ }\textbf {\bibinfo {volume} {342}},\ \bibinfo {pages} {604} (\bibinfo {year} {2013})}\BibitemShut {NoStop}%
\bibitem [{\citenamefont {Transtrum}\ \emph {et~al.}(2015)\citenamefont {Transtrum}, \citenamefont {Machta}, \citenamefont {Brown}, \citenamefont {Daniels}, \citenamefont {Myers},\ and\ \citenamefont {Sethna}}]{transtrum2015Perspective}%
  \BibitemOpen
  \bibfield  {author} {\bibinfo {author} {\bibfnamefont {M.~K.}\ \bibnamefont {Transtrum}}, \bibinfo {author} {\bibfnamefont {B.~B.}\ \bibnamefont {Machta}}, \bibinfo {author} {\bibfnamefont {K.~S.}\ \bibnamefont {Brown}}, \bibinfo {author} {\bibfnamefont {B.~C.}\ \bibnamefont {Daniels}}, \bibinfo {author} {\bibfnamefont {C.~R.}\ \bibnamefont {Myers}},\ and\ \bibinfo {author} {\bibfnamefont {J.~P.}\ \bibnamefont {Sethna}},\ }\bibfield  {title} {\bibinfo {title} {Perspective: {{Sloppiness}} and emergent theories in physics, biology, and beyond},\ }\href {https://doi.org/10.1063/1.4923066} {\bibfield  {journal} {\bibinfo  {journal} {The Journal of Chemical Physics}\ }\textbf {\bibinfo {volume} {143}},\ \bibinfo {pages} {010901} (\bibinfo {year} {2015})}\BibitemShut {NoStop}%
\bibitem [{\citenamefont {Chen}(2025)}]{chen2025ising2D}%
  \BibitemOpen
  \bibfield  {author} {\bibinfo {author} {\bibfnamefont {Q.}~\bibnamefont {Chen}},\ }\href {https://doi.org/10.5281/zenodo.17118297} {\bibinfo {title} {qianyangchen/isingthermodynamicefficiency: v1.0.1}} (\bibinfo {year} {2025}),\ \bibinfo {note} {software version v1.0.1}\BibitemShut {NoStop}%
\bibitem [{\citenamefont {Kleidon}\ \emph {et~al.}(2024)\citenamefont {Kleidon}, \citenamefont {Gozzi}, \citenamefont {Buccianti},\ and\ \citenamefont {Sauro~Graziano}}]{KLEIDON2024173409}%
  \BibitemOpen
  \bibfield  {author} {\bibinfo {author} {\bibfnamefont {A.}~\bibnamefont {Kleidon}}, \bibinfo {author} {\bibfnamefont {C.}~\bibnamefont {Gozzi}}, \bibinfo {author} {\bibfnamefont {A.}~\bibnamefont {Buccianti}},\ and\ \bibinfo {author} {\bibfnamefont {R.}~\bibnamefont {Sauro~Graziano}},\ }\bibfield  {title} {\bibinfo {title} {Type of probability distribution reflects how close mixing dynamics in river chemistry are to thermodynamic equilibrium},\ }\href {https://doi.org/10.1016/j.scitotenv.2024.173409} {\bibfield  {journal} {\bibinfo  {journal} {Science of The Total Environment}\ }\textbf {\bibinfo {volume} {941}},\ \bibinfo {pages} {173409} (\bibinfo {year} {2024})}\BibitemShut {NoStop}%
\bibitem [{\citenamefont {Beck}(2007)}]{beck2007Statistics}%
  \BibitemOpen
  \bibfield  {author} {\bibinfo {author} {\bibfnamefont {C.}~\bibnamefont {Beck}},\ }\bibfield  {title} {\bibinfo {title} {Statistics of {{Three-Dimensional Lagrangian Turbulence}}},\ }\href {https://doi.org/10.1103/PhysRevLett.98.064502} {\bibfield  {journal} {\bibinfo  {journal} {Physical Review Letters}\ }\textbf {\bibinfo {volume} {98}},\ \bibinfo {pages} {064502} (\bibinfo {year} {2007})}\BibitemShut {NoStop}%
\bibitem [{\citenamefont {Gustavsson}\ and\ \citenamefont {Mehlig}(2016)}]{gustavsson2016Statistical}%
  \BibitemOpen
  \bibfield  {author} {\bibinfo {author} {\bibfnamefont {K.}~\bibnamefont {Gustavsson}}\ and\ \bibinfo {author} {\bibfnamefont {B.}~\bibnamefont {Mehlig}},\ }\bibfield  {title} {\bibinfo {title} {Statistical models for spatial patterns of heavy particles in turbulence},\ }\href {https://doi.org/10.1080/00018732.2016.1164490} {\bibfield  {journal} {\bibinfo  {journal} {Advances in Physics}\ }\textbf {\bibinfo {volume} {65}},\ \bibinfo {pages} {1} (\bibinfo {year} {2016})}\BibitemShut {NoStop}%
\bibitem [{\citenamefont {Bialek}\ \emph {et~al.}(2012)\citenamefont {Bialek}, \citenamefont {Cavagna}, \citenamefont {Giardina}, \citenamefont {Mora}, \citenamefont {Silvestri}, \citenamefont {Viale},\ and\ \citenamefont {Walczak}}]{bialek2012Statistical}%
  \BibitemOpen
  \bibfield  {author} {\bibinfo {author} {\bibfnamefont {W.}~\bibnamefont {Bialek}}, \bibinfo {author} {\bibfnamefont {A.}~\bibnamefont {Cavagna}}, \bibinfo {author} {\bibfnamefont {I.}~\bibnamefont {Giardina}}, \bibinfo {author} {\bibfnamefont {T.}~\bibnamefont {Mora}}, \bibinfo {author} {\bibfnamefont {E.}~\bibnamefont {Silvestri}}, \bibinfo {author} {\bibfnamefont {M.}~\bibnamefont {Viale}},\ and\ \bibinfo {author} {\bibfnamefont {A.~M.}\ \bibnamefont {Walczak}},\ }\bibfield  {title} {\bibinfo {title} {Statistical mechanics for natural flocks of birds},\ }\href {https://doi.org/10.1073/pnas.1118633109} {\bibfield  {journal} {\bibinfo  {journal} {Proceedings of the National Academy of Sciences}\ }\textbf {\bibinfo {volume} {109}},\ \bibinfo {pages} {4786} (\bibinfo {year} {2012})}\BibitemShut {NoStop}%
\bibitem [{\citenamefont {Nghiem}\ \emph {et~al.}(2018)\citenamefont {Nghiem}, \citenamefont {Telenczuk}, \citenamefont {Marre}, \citenamefont {Destexhe},\ and\ \citenamefont {Ferrari}}]{nghiem2018Maximumentropy}%
  \BibitemOpen
  \bibfield  {author} {\bibinfo {author} {\bibfnamefont {T.-A.}\ \bibnamefont {Nghiem}}, \bibinfo {author} {\bibfnamefont {B.}~\bibnamefont {Telenczuk}}, \bibinfo {author} {\bibfnamefont {O.}~\bibnamefont {Marre}}, \bibinfo {author} {\bibfnamefont {A.}~\bibnamefont {Destexhe}},\ and\ \bibinfo {author} {\bibfnamefont {U.}~\bibnamefont {Ferrari}},\ }\bibfield  {title} {\bibinfo {title} {Maximum-entropy models reveal the excitatory and inhibitory correlation structures in cortical neuronal activity},\ }\href {https://doi.org/10.1103/PhysRevE.98.012402} {\bibfield  {journal} {\bibinfo  {journal} {Physical Review E}\ }\textbf {\bibinfo {volume} {98}},\ \bibinfo {pages} {012402} (\bibinfo {year} {2018})}\BibitemShut {NoStop}%
\bibitem [{\citenamefont {Zanoci}\ \emph {et~al.}(2019)\citenamefont {Zanoci}, \citenamefont {Dehghani},\ and\ \citenamefont {Tegmark}}]{zanoci2019Ensemble}%
  \BibitemOpen
  \bibfield  {author} {\bibinfo {author} {\bibfnamefont {C.}~\bibnamefont {Zanoci}}, \bibinfo {author} {\bibfnamefont {N.}~\bibnamefont {Dehghani}},\ and\ \bibinfo {author} {\bibfnamefont {M.}~\bibnamefont {Tegmark}},\ }\bibfield  {title} {\bibinfo {title} {Ensemble inhibition and excitation in the human cortex: {{An Ising-model}} analysis with uncertainties},\ }\href {https://doi.org/10.1103/PhysRevE.99.032408} {\bibfield  {journal} {\bibinfo  {journal} {Physical Review E}\ }\textbf {\bibinfo {volume} {99}},\ \bibinfo {pages} {032408} (\bibinfo {year} {2019})}\BibitemShut {NoStop}%
\bibitem [{\citenamefont {Hunter}(2007)}]{hunter2007Curved}%
  \BibitemOpen
  \bibfield  {author} {\bibinfo {author} {\bibfnamefont {D.~R.}\ \bibnamefont {Hunter}},\ }\bibfield  {title} {\bibinfo {title} {Curved exponential family models for social networks},\ }\href {https://doi.org/10.1016/j.socnet.2006.08.005} {\bibfield  {journal} {\bibinfo  {journal} {Social Networks}\ }\textbf {\bibinfo {volume} {29}},\ \bibinfo {pages} {216} (\bibinfo {year} {2007})}\BibitemShut {NoStop}%
\bibitem [{\citenamefont {Mantegna}\ and\ \citenamefont {Stanley}(1995)}]{mantegna1995Scaling}%
  \BibitemOpen
  \bibfield  {author} {\bibinfo {author} {\bibfnamefont {R.~N.}\ \bibnamefont {Mantegna}}\ and\ \bibinfo {author} {\bibfnamefont {H.~E.}\ \bibnamefont {Stanley}},\ }\bibfield  {title} {\bibinfo {title} {Scaling behaviour in the dynamics of an economic index},\ }\href {https://doi.org/10.1038/376046a0} {\bibfield  {journal} {\bibinfo  {journal} {Nature}\ }\textbf {\bibinfo {volume} {376}},\ \bibinfo {pages} {46} (\bibinfo {year} {1995})}\BibitemShut {NoStop}%
\bibitem [{\citenamefont {Jiang}\ and\ \citenamefont {Zhou}(2010)}]{jiang2010Complex}%
  \BibitemOpen
  \bibfield  {author} {\bibinfo {author} {\bibfnamefont {Z.-Q.}\ \bibnamefont {Jiang}}\ and\ \bibinfo {author} {\bibfnamefont {W.-X.}\ \bibnamefont {Zhou}},\ }\bibfield  {title} {\bibinfo {title} {Complex stock trading network among investors},\ }\href {https://doi.org/10.1016/j.physa.2010.07.024} {\bibfield  {journal} {\bibinfo  {journal} {Physica A: Statistical Mechanics and its Applications}\ }\textbf {\bibinfo {volume} {389}},\ \bibinfo {pages} {4929} (\bibinfo {year} {2010})}\BibitemShut {NoStop}%
\bibitem [{\citenamefont {Marsili}\ and\ \citenamefont {Zhang}(1998)}]{marsili1998Interacting}%
  \BibitemOpen
  \bibfield  {author} {\bibinfo {author} {\bibfnamefont {M.}~\bibnamefont {Marsili}}\ and\ \bibinfo {author} {\bibfnamefont {Y.-C.}\ \bibnamefont {Zhang}},\ }\bibfield  {title} {\bibinfo {title} {Interacting {{Individuals Leading}} to {{Zipf}}'s {{Law}}},\ }\href {https://doi.org/10.1103/PhysRevLett.80.2741} {\bibfield  {journal} {\bibinfo  {journal} {Physical Review Letters}\ }\textbf {\bibinfo {volume} {80}},\ \bibinfo {pages} {2741} (\bibinfo {year} {1998})}\BibitemShut {NoStop}%
\bibitem [{\citenamefont {Barab{\'a}si}\ and\ \citenamefont {Albert}(1999)}]{barabasi1999Emergence}%
  \BibitemOpen
  \bibfield  {author} {\bibinfo {author} {\bibfnamefont {A.-L.}\ \bibnamefont {Barab{\'a}si}}\ and\ \bibinfo {author} {\bibfnamefont {R.}~\bibnamefont {Albert}},\ }\bibfield  {title} {\bibinfo {title} {Emergence of {{Scaling}} in {{Random Networks}}},\ }\href {https://doi.org/10.1126/science.286.5439.509} {\bibfield  {journal} {\bibinfo  {journal} {Science}\ }\textbf {\bibinfo {volume} {286}},\ \bibinfo {pages} {509} (\bibinfo {year} {1999})}\BibitemShut {NoStop}%
\bibitem [{\citenamefont {Newman}(2001)}]{newman2001Clustering}%
  \BibitemOpen
  \bibfield  {author} {\bibinfo {author} {\bibfnamefont {M.~E.~J.}\ \bibnamefont {Newman}},\ }\bibfield  {title} {\bibinfo {title} {Clustering and preferential attachment in growing networks},\ }\href {https://doi.org/10.1103/PhysRevE.64.025102} {\bibfield  {journal} {\bibinfo  {journal} {Physical Review E}\ }\textbf {\bibinfo {volume} {64}},\ \bibinfo {pages} {025102} (\bibinfo {year} {2001})}\BibitemShut {NoStop}%
\bibitem [{\citenamefont {Barab{\'a}si}(2005)}]{barabasi2005Origin}%
  \BibitemOpen
  \bibfield  {author} {\bibinfo {author} {\bibfnamefont {A.-L.}\ \bibnamefont {Barab{\'a}si}},\ }\bibfield  {title} {\bibinfo {title} {The origin of bursts and heavy tails in human dynamics},\ }\href {https://doi.org/10.1038/nature03459} {\bibfield  {journal} {\bibinfo  {journal} {Nature}\ }\textbf {\bibinfo {volume} {435}},\ \bibinfo {pages} {207} (\bibinfo {year} {2005})}\BibitemShut {NoStop}%
\bibitem [{\citenamefont {Muchnik}\ \emph {et~al.}(2013)\citenamefont {Muchnik}, \citenamefont {Pei}, \citenamefont {Parra}, \citenamefont {Reis}, \citenamefont {Andrade~Jr}, \citenamefont {Havlin},\ and\ \citenamefont {Makse}}]{muchnik2013Origins}%
  \BibitemOpen
  \bibfield  {author} {\bibinfo {author} {\bibfnamefont {L.}~\bibnamefont {Muchnik}}, \bibinfo {author} {\bibfnamefont {S.}~\bibnamefont {Pei}}, \bibinfo {author} {\bibfnamefont {L.~C.}\ \bibnamefont {Parra}}, \bibinfo {author} {\bibfnamefont {S.~D.~S.}\ \bibnamefont {Reis}}, \bibinfo {author} {\bibfnamefont {J.~S.}\ \bibnamefont {Andrade~Jr}}, \bibinfo {author} {\bibfnamefont {S.}~\bibnamefont {Havlin}},\ and\ \bibinfo {author} {\bibfnamefont {H.~A.}\ \bibnamefont {Makse}},\ }\bibfield  {title} {\bibinfo {title} {Origins of power-law degree distribution in the heterogeneity of human activity in social networks},\ }\href {https://doi.org/10.1038/srep01783} {\bibfield  {journal} {\bibinfo  {journal} {Scientific Reports}\ }\textbf {\bibinfo {volume} {3}},\ \bibinfo {pages} {1783} (\bibinfo {year} {2013})}\BibitemShut {NoStop}%
\bibitem [{\citenamefont {Niven}\ and\ \citenamefont {Andresen}(2010)}]{niven2010Jaynes}%
  \BibitemOpen
  \bibfield  {author} {\bibinfo {author} {\bibfnamefont {R.~K.}\ \bibnamefont {Niven}}\ and\ \bibinfo {author} {\bibfnamefont {B.}~\bibnamefont {Andresen}},\ }\bibfield  {title} {\bibinfo {title} {Jaynes' {{Maximum Entropy Principle}}, {{Riemannian Metrics}} and {{Generalised Least Action Bound}}},\ }in\ \href {https://doi.org/10.1142/9789814277327_0008} {\emph {\bibinfo {booktitle} {Complex {{Physical}}, {{Biophysical}} and {{Econophysical Systems}}}}}\ (\bibinfo  {publisher} {World Scientific},\ \bibinfo {address} {The Australian National University, Canberra},\ \bibinfo {year} {2010})\ pp.\ \bibinfo {pages} {283--317}\BibitemShut {NoStop}%
\bibitem [{\citenamefont {{de Groot}}\ and\ \citenamefont {Mazur}(2013)}]{degroot2013Nonequilibrium}%
  \BibitemOpen
  \bibfield  {author} {\bibinfo {author} {\bibfnamefont {S.~R.}\ \bibnamefont {{de Groot}}}\ and\ \bibinfo {author} {\bibfnamefont {P.}~\bibnamefont {Mazur}},\ }\href@noop {} {\emph {\bibinfo {title} {Non-Equilibrium Thermodynamics}}},\ Dover {{Books}} on {{Physics}}\ (\bibinfo  {publisher} {Dover Publications},\ \bibinfo {address} {Newburyport},\ \bibinfo {year} {2013})\BibitemShut {NoStop}%
\bibitem [{\citenamefont {Blahut}(2003)}]{blahut2003computation}%
  \BibitemOpen
  \bibfield  {author} {\bibinfo {author} {\bibfnamefont {R.}~\bibnamefont {Blahut}},\ }\bibfield  {title} {\bibinfo {title} {Computation of channel capacity and rate-distortion functions},\ }\href {https://doi.org/10.1109/TIT.1972.1054855} {\bibfield  {journal} {\bibinfo  {journal} {IEEE transactions on Information Theory}\ }\textbf {\bibinfo {volume} {18}},\ \bibinfo {pages} {460} (\bibinfo {year} {2003})}\BibitemShut {NoStop}%
\bibitem [{\citenamefont {Wilson}(2008)}]{wilson2008boltzmann}%
  \BibitemOpen
  \bibfield  {author} {\bibinfo {author} {\bibfnamefont {A.}~\bibnamefont {Wilson}},\ }\bibfield  {title} {\bibinfo {title} {{Boltzmann, Lotka and Volterra and spatial structural evolution: an integrated methodology for some dynamical systems}},\ }\href {https://doi.org/10.1098/rsif.2007.1288} {\bibfield  {journal} {\bibinfo  {journal} {Journal of The Royal Society Interface}\ }\textbf {\bibinfo {volume} {5}},\ \bibinfo {pages} {865} (\bibinfo {year} {2008})}\BibitemShut {NoStop}%
\bibitem [{\citenamefont {Salge}\ \emph {et~al.}(2015)\citenamefont {Salge}, \citenamefont {Ay}, \citenamefont {Polani},\ and\ \citenamefont {Prokopenko}}]{salge2015zipf}%
  \BibitemOpen
  \bibfield  {author} {\bibinfo {author} {\bibfnamefont {C.}~\bibnamefont {Salge}}, \bibinfo {author} {\bibfnamefont {N.}~\bibnamefont {Ay}}, \bibinfo {author} {\bibfnamefont {D.}~\bibnamefont {Polani}},\ and\ \bibinfo {author} {\bibfnamefont {M.}~\bibnamefont {Prokopenko}},\ }\bibfield  {title} {\bibinfo {title} {Zipf’s law: balancing signal usage cost and communication efficiency},\ }\href {https://doi.org/10.1371/journal.pone.0139475} {\bibfield  {journal} {\bibinfo  {journal} {PLoS One}\ }\textbf {\bibinfo {volume} {10}},\ \bibinfo {pages} {e0139475} (\bibinfo {year} {2015})}\BibitemShut {NoStop}%
\bibitem [{\citenamefont {Brody}\ and\ \citenamefont {Rivier}(1995)}]{brody1995Geometrical}%
  \BibitemOpen
  \bibfield  {author} {\bibinfo {author} {\bibfnamefont {D.}~\bibnamefont {Brody}}\ and\ \bibinfo {author} {\bibfnamefont {N.}~\bibnamefont {Rivier}},\ }\bibfield  {title} {\bibinfo {title} {Geometrical aspects of statistical mechanics},\ }\href {https://doi.org/10.1103/PhysRevE.51.1006} {\bibfield  {journal} {\bibinfo  {journal} {Physical Review E}\ }\textbf {\bibinfo {volume} {51}},\ \bibinfo {pages} {1006} (\bibinfo {year} {1995})}\BibitemShut {NoStop}%
\bibitem [{\citenamefont {Metropolis}\ \emph {et~al.}(1953)\citenamefont {Metropolis}, \citenamefont {Rosenbluth}, \citenamefont {Rosenbluth}, \citenamefont {Teller},\ and\ \citenamefont {Teller}}]{metropolis1953Equation}%
  \BibitemOpen
  \bibfield  {author} {\bibinfo {author} {\bibfnamefont {N.}~\bibnamefont {Metropolis}}, \bibinfo {author} {\bibfnamefont {A.~W.}\ \bibnamefont {Rosenbluth}}, \bibinfo {author} {\bibfnamefont {M.~N.}\ \bibnamefont {Rosenbluth}}, \bibinfo {author} {\bibfnamefont {A.~H.}\ \bibnamefont {Teller}},\ and\ \bibinfo {author} {\bibfnamefont {E.}~\bibnamefont {Teller}},\ }\bibfield  {title} {\bibinfo {title} {Equation of {{State Calculations}} by {{Fast Computing Machines}}},\ }\href {https://doi.org/10.1063/1.1699114} {\bibfield  {journal} {\bibinfo  {journal} {The Journal of Chemical Physics}\ }\textbf {\bibinfo {volume} {21}},\ \bibinfo {pages} {1087} (\bibinfo {year} {1953})}\BibitemShut {NoStop}%
\bibitem [{\citenamefont {Kikuchi}(1951)}]{kikuchi1951Theory}%
  \BibitemOpen
  \bibfield  {author} {\bibinfo {author} {\bibfnamefont {R.}~\bibnamefont {Kikuchi}},\ }\bibfield  {title} {\bibinfo {title} {A {{Theory}} of {{Cooperative Phenomena}}},\ }\href {https://doi.org/10.1103/PhysRev.81.988} {\bibfield  {journal} {\bibinfo  {journal} {Physical Review}\ }\textbf {\bibinfo {volume} {81}},\ \bibinfo {pages} {988} (\bibinfo {year} {1951})}\BibitemShut {NoStop}%
\end{thebibliography}
\end{document}